\documentclass[adp,fleqn]{w-art}
\usepackage{times}
\usepackage{w-thm}
\usepackage{braket}
\usepackage{mathrsfs}
\usepackage[]{graphicx}
\chardef\bslash=`\\ 

\newcommand{\tj}[6]{\left(\begin{array}{ccc} #1&#3&#5\\ #2&#4&#6\end{array}\right)}

\newcommand{\MF}{CH$_3$F}

\newcommand{\Mn}{Mn$_{12}$}
\newcommand{\Mnac}{Mn$_{12\mathrm{ac}}$}
\newcommand{\Fe}{Fe$_8$}
\newcommand{\LiCs}{{$^{6}$Li$^{133}$Cs}}
\newcommand{\NaK}{$^{23}$Na$^{40}$K}
\newcommand{\RbCs}{$^{87}$Rb$^{133}$Cs}
\newcommand{\KRb}{$^{40}$K$^{87}$Rb}
\newcommand{\LiNa}{{$^{6}$Li$^{23}$Na}}

\newcommand \beq{\begin{eqnarray}}
\newcommand \eeq{\end{eqnarray}}
\def\Jvol<#1,#2,#3>{#1}
\def\Jpage<#1,#2,#3>{#2}
\def\Jyear<#1,#2,#3>{#3}
\newcommand\journal[1]{\textbf{\Jvol<#1>}, \Jpage<#1> (\Jyear<#1>)}

\newcommand\PRL[1]{Phys.\ Rev.\ Lett.\ \journal{#1}}

\newcommand\PRA[1]{Phys.\ Rev.\ A \journal{#1}}

\newcommand\PREP[1]{Phys.\ Rep.\ \journal{#1}}

\begin{document}
\DOIsuffix{theDOIsuffix}
\Volume{12}
\Issue{1}
\Copyrightissue{01}
\Month{01}
\Year{2003}
\pagespan{1}{}
\Receiveddate{15 November 1900}
\Accepteddate{2 December 1900}
\keywords{quantum simulation, quantum many body physics, cold molecules, symmetric top molecules, magnetic atoms, linear Stark effect, Einstein-de Haas effect,
methyl fluoride, molecular magnets \\}


\title{Simulating quantum magnets with symmetric top molecules}

\author[Wall]{Michael L Wall\footnote{Corresponding
     author \quad E-mail: {\sf mwall.physics@gmail.com}}} 
\address
{Department of Physics, Colorado School of Mines, Golden, Colorado 80401, USA }
\author[Maeda]{Kenji Maeda\footnote{E-mail: {\sf kenji.bosefermi@gmail.com}}}
\author[Carr]{Lincoln D Carr}

\begin{abstract}
We establish a correspondence between the electric dipole matrix elements of a polyatomic symmetric top molecule in a state with nonzero projection of the total angular momentum on the symmetry axis of the molecule and the magnetic dipole matrix elements of a magnetic dipole associated with an elemental spin $F$. 
It is shown that this correspondence makes it possible to 
perform quantum simulation of the single-particle spectrum and the dipole-dipole interactions of magnetic dipoles in a static external magnetic field $\bf{B}$ 
with symmetric top molecules subject to a static external electric field $\bf{E}_{\mathrm{DC}}$.  We further show that no such correspondence exists for $^1\Sigma$ molecules in static fields, such as the alkali metal dimers.  The effective spin angular momentum of the simulated magnetic dipole corresponds to the rotational angular momentum of the symmetric top molecule, 
and so quantum simulation of arbitrarily large integer spins is possible.  
Further, taking the molecule CH$_3$F as an example, we show that the characteristic dipole-dipole interaction energies of the simulated magnetic dipole 
are a factor of 620, 600, and 310 larger than for the highly magnetic atoms Chromium, Erbium, and Dysprosium, respectively.  
We present several applications of our correspondence for many-body physics, 
including long-range and anisotropic spin models with arbitrary integer spin $S$ using symmetric top molecules in optical lattices, 
quantum simulation of molecular magnets, and spontaneous demagnetization of Bose-Einstein condensates due to dipole-dipole interactions.  Our results are expected to be relevant as cold symmetric top molecules reach quantum degeneracy through Stark deceleration and opto-electrical cooling.
\end{abstract}

\maketitle

\section{Introduction}
\label{sect1}

The realization that the phenomena of electricity and magnetism are related manifestations of the same field is a landmark in the history of physics, 
marking the first successful unification of theories.  
In spite of the fact that electricity and magnetism are derived from the same physical source, electric phenomena and magnetic phenomena are not identical.  
This can be seen classically by the fact that the electric field is a vector, transforming into its mirror image under inversion of all coordinates, 
while the magnetic field is a pseudo-vector, transforming into minus its mirror image under inversion of all coordinates.  
In quantum mechanics, this difference between electric and magnetic fields has the consequence 
that matrix elements of the electric dipole operator vanish unless the two states have opposite parity, 
while magnetic dipole matrix elements connect states which have the same parity.  
Hence, an elemental object can possess a magnetic dipole but not an electric dipole, 
as possession of an electric dipole would violate both parity and time-reversal symmetry\footnote{As discussed also in Sec.~\ref{sec:Hfs}, parity and time-reversal are not exact symmetries of the universe within the understanding of the standard model.  However, their violations are very small.  Ultracold molecules, incidentally, provide a highly accurate platform to set bounds on these violations~\cite{DeMilleEDM,electronEDM,cornellEDM}.}.

Even in light of the fundamental differences between electric and magnetic dipoles, interactions between two electric or magnetic dipoles 
are described by a Hamiltonian of the same form, namely~{\cite{Ueda_2010}}
\beq
\label{eq:HDDI1}\hat{H}_{\mathrm{DDI};\mathcal{D}}&=&
\frac{C_{\mathcal{D}}}{4\pi}\frac{\hat{\boldsymbol{\mathcal{D}}}_1\cdot\hat{\boldsymbol{\mathcal{D}}}_2
-3\left(\hat{\boldsymbol{\mathcal{D}}}_1\cdot \mathbf{e}_r\right)\left(\mathbf{e}_r\cdot \hat{\boldsymbol{\mathcal{D}}}_2\right)}{r^3}\, ,
\eeq
where $\hat{\boldsymbol{\mathcal{D}}}_i$ is the dipole moment operator $\hat{\boldsymbol{\mathcal{D}}}$ of the $i^{\mathrm{th}}$ particle 
with $\hat{\boldsymbol{\mathcal{D}}}=\hat{\bf{d}},\:\hat{\boldsymbol{\mu}}$ for electric and magnetic dipoles, respectively.  
Also, $r$ is the distance between the two dipoles, $\mathbf{e}_r$ is a unit vector pointing from dipole 1 to dipole 2, 
and the coupling {coefficients} for the electric and magnetic dipole-dipole interactions are given by
\beq
\label{eq:couplingconstants}C_{d}&=&1/\varepsilon_0\, ,\quad 
C_{\mu}~=~\mu_0\, ,
\eeq
with $\varepsilon_0$ the vacuum permittivity and $\mu_0$ the vacuum permeability.  
The dipole-dipole interaction Eq.~(\ref{eq:HDDI1}) is of interest from the point of view of many-body physics because it is long-range and anisotropic.  
The differences in coupling {coefficients}, Eq.~(\ref{eq:couplingconstants}), demonstrate a practical difference between electric and magnetic dipoles.  
Namely, the interactions between electric dipoles on typical atomic scales are roughly $10^4$ times larger than for magnetic dipoles.  
Hence, {quantum simulation of a gas} of magnetic dipoles with electric dipoles would provide 
access to the physics of magnetic dipoles in more strongly correlated regimes, on shorter timescales, and at lower density 
than for the true magnetic dipoles encountered {in nature}.  In this work, we show that such a quantum simulation is indeed possible.

\begin{figure}[tbp]
\centerline{\includegraphics[width=0.75\columnwidth]{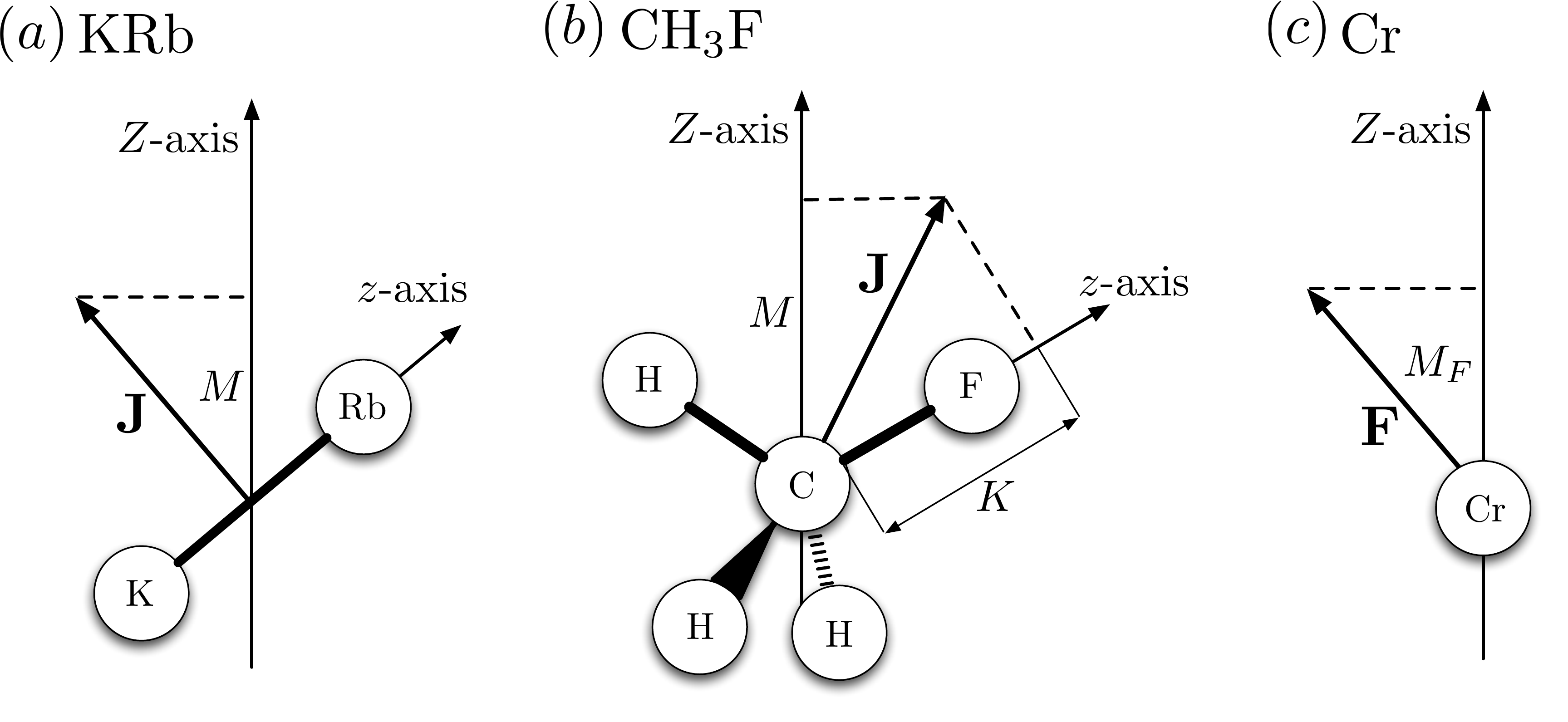}}
\caption{\label{fig:Schematic}  
Schematic of {angular momentum geometry} for (a) KRb, a linear rigid rotor; (b) {\MF}, a polyatomic symmetric top molecule; and (c) Cr, a highly magnetic atom.  Here, we display both the space-fixed $Z$-axis and the molecule-fixed $z$-axis as well as the projections of the angular momentum along them, see the discussion following Eq.~(\ref{eq:Hrot}).  In the molecular cases, panels (a) and (b), the total angular momentum consists only of rotation.  {For the linear rigid rotor in panel (a) this angular momentum has no projection on the molecular symmetry axis, but for the symmetric top molecule in panel (b) the angular momentum can have some projection $K$ on this axis.}
}
\end{figure}

It is only in the last ten years that quantum degenerate gases with significant dipole-dipole interactions have been an experimental reality.  
The first {strongly dipolar gas} to reach quantum degeneracy was Chromium~\cite{Griesmaier_Werner_05}, 
which {has} a magnetic dipole moment of $6\mu_B$, where $\mu_B$ is the Bohr magneton.  A clear manifestation of the effects of dipole-dipole interactions were seen in the $d$-wave symmetry of the atomic cloud expansion following collapse of a Chromium Bose-Einstein condensate~\cite{Collapse}.  Even for the alkali metals with magnetic dipole moments of $1\mu_B$, dipolar effects have been seen in some cases~\cite{Vengalattore_2008,Fattori}.
For Chromium, the ratio of the dipole-dipole interaction energy to the interaction energy due to isotropic collisions is $0.15$, 
and so the effect of dipole-dipole interactions in this system is considered to be a {small perturbation}.  
{Recently}, degenerate gases of Erbium~\cite{Er} and both bosonic~\cite{DyBEC} and fermionic~\cite{DyFermi} Dysprosium have been produced, 
which have magnetic dipole moments of $7\mu_B$ and $10\mu_B$, respectively.  Assuming that the scattering lengths of Erbium and Dysprosium are both roughly 100 Bohr radii, as is the case for Chromium, this gives ratios of the dipole-dipole interaction energy to the isotropic interaction energy of 0.67 and 1.33, respectively.  Significant progress has also been made on the production of ultracold polar molecules~\cite{MolrevNJP}, 
with KRb~\cite{KRb} being the first to reach the ultracold regime.  
Several other species are currently being investigated~\cite{Quemener_Julienne_12}, 
the majority of which are diatomic molecules consisting of two alkali metal atoms~\cite{RbCs,LiCs,NaK,LiNa}.  
The alkali metal dimers are $^1\Sigma$ molecules, having no electronic spin or orbital angular momentum, 
and so their angular momentum structure corresponds to that in Fig.~\ref{fig:Schematic}{(a)}.  {Namely, the total angular momentum $\hat{\mathbf{J}}$ consists only of rotation, 
and is perpendicular to the internuclear axis. }

In many theoretical works on the many-body physics of dipoles, it is presumed that only the magnitude of the dipole moment has importance.  In particular, it is supposed that the origin of the dipole moment, whether it is magnetic or electric, is of no importance.  For the case of polarized dipoles, in which only a single quantum state with a resonant dipole moment is populated, this presumption is indeed true, as we discuss in Sec.~\ref{sec:polarized}.  However, for the multi-component case, the structure of the dipole matrix elements within the allowed state manifold and the energies of these states are often very different depending on the origin of the dipole moments.  In particular, while the alkali metal dimers typically have characteristic dipole-dipole interaction energies 
several orders of magnitude larger than for atoms with magnetic dipole moments, 
the energies and expected dipole moments of these molecules are very different from magnetic atoms.  
Magnetic dipoles associated with an elemental angular momentum {$\hat{\bf F}$} display a linear coupling to an externally applied magnetic field 
such that {differences in energies} 
between states with projections of {$\hat{\bf F}$} along the field direction of $M_F$ and $M_{F}\pm 1$ are {the same}. 
In addition, the magnetic dipole moment behaves as though it points along {$\hat{\bf F}$}, 
and so the matrix elements of the magnetic dipole moment along any direction in space can be non-vanishing.  
Loosely speaking, such a magnetic dipole moment is ``always on."  On the other hand, the electric dipole moment of a $^1\Sigma$ points along the {inter-nuclear} axis, 
and {its} expectation vanishes along any direction in space due to effective averaging by the rotation of the molecule.  
A large static electric field must be applied in order to polarize the dipole moment along a particular direction in space.  
However, as the coupling of a $^1\Sigma$ molecule to an external electric field occurs only in second and higher orders of perturbation theory, 
the energy levels and dipole moments do not resemble those of a magnetic dipole.  Thus experiments like those of Bruno Laburthe-Tolra on the many body physics of Chromium atoms in optical lattices~\cite{Pasquiou_Bismut_81,Pasquiou_Bismut_106,Pasquiou_Marechal_108,dePaz_Chotia_12} cannot be reproduced or explored in the many $^1\Sigma$ molecules now under intensive development in laboratories all over the world~\cite{Quemener_Julienne_12,RbCs,LiCs,NaK,LiNa}.

{We will overcome these difficulties in quantum simulation of magnetic dipoles with $^1\Sigma$ molecules by focusing on symmetric top molecules.}  
A symmetric top molecule is a polyatomic molecule with cylindrical symmetry, resulting in a doubly degenerate eigenvalue of the inertia tensor.  
This implies that the angular momentum of a symmetric top molecule can have some component in {the direction of the symmetry axis}, 
see Fig.~\ref{fig:Schematic}{(b)}.  
In particular, when the projection of the angular momentum along the molecular symmetry axis is non-vanishing, 
a symmetric top molecule displays a linear response to an externally applied electric field, 
{so that} the energies and dipole matrix elements mimic those of an elemental magnetic dipole.  We remark that a linear response to an external electric field is not indicative of a permanent electric dipole moment in the space-fixed frame, as is discussed further in Sec.~\ref{sec:Hfs}.  
In addition to their fundamental interest in quantum simulation given in this paper, 
symmetric top molecules are also promising candidates for reaching quantum degeneracy precisely because of their sensitivity to electric fields.  
This sensitivity allows for molecules to be slowed by the application of an electric field gradient known as a Stark decelerator~\cite{Bethlem_Berden_99}.  
In addition, opto-electrical cooling, {where} electric field interaction energy is used rather than photon recoil to remove energy from translational motion, 
is now a demonstrated technology for the symmetric top molecule {\MF}~\cite{CH3FOEC}.  For this reason we take {\MF} as the principle physical example of a symmetric top molecule for this article.  To date, the most successful means of producing ultracold molecules have been assembly of ultracold molecules from ultracold atoms by magnetic~\cite{KRb} or optical~\cite{Ulmanis} means or direct laser cooling of molecules~\cite{SrF}, all of which work only for specific species.  In contrast, opto-electrical cooling has the potential to work for any symmetric top molecule, of which {\MF} represents an initial experimental demonstration.  Opto-electrical cooling hence opens up a new paradigm for the production of ultracold quantum gases, especially of common molecules relevant to chemistry.

The ability to mimic the single-particle and two-particle matrix elements of quantum magnets with symmetric top molecules also implies the ability to engineer quantum simulations of dilute gases or crystals of quantum magnets with symmetric top molecules.  These quantum simulations open up a wealth of many-body physics, presented in Sec.~\ref{sec:applications}, including new regimes of many-body physics which are not accessible with naturally occurring magnetic dipoles.  For example, we show that polarized gases of symmetric top molecules in a single quantum state interact much more strongly than single-component gases of magnetic dipoles, and require much smaller electric fields than $^1\Sigma$ molecules.  Long-range and anisotropic lattice spin models can be simulated with symmetric top molecules in deep optical lattices.  In addition to the long-range version of the XXZ model where simulation is also possible with linear rigid rotors~\cite{Gorshkov_Manmana_11,Gorshkov_Manmana_11b}, we demonstrate how to simulate spin models which do not conserve magnetization, a feat which requires fine-tuning of a large number of fields to achieve with linear rigid rotors~\cite{Gorshkov_Hazzard_13}.  A major advantage of simulations with symmetric top molecules is that any effective spin may be chosen without fine-tuning of fields by populating a different rotational manifold.  Furthermore, the exchange statistics of the molecule is not associated with this effective spin, and so symmetric top molecules provide a quantum simulation of a fermionic particle with a large integer spin, something which does not occur for actual magnetic dipoles.  Using the connection of large-spin physics, we discuss using symmetric top molecules to simulate molecular magnets in condensed matter.  Finally, we discuss simulations of the spontaneous demagnetization of degenerate Bose gases such as has been seen for Chromium~\cite{Pasquiou_Marechal_108,dePaz_Chotia_12}.  Here, simulations with symmetric top molecules not only offer the advantages of shorter timescales for dynamics and lower requirements for density, but also allow the possibility of exploring new strongly correlated regimes which would not be accessible to any magnetic atom.

This paper is outlined as follows.  
In Sec.~\ref{sect2} we review the rotational structure of symmetric top molecules and discuss the coupling of symmetric top molecules to external fields.  
In Sec.~\ref{sec:Reln}, we {review} the structure of magnetic dipoles and 
present a correspondence between the energies and dipole matrix elements of symmetric top molecules and magnetic dipoles.  
Sec.~\ref{sec:mapping}, in particular Table~\ref{table:mapping}, contains {such} correspondence, which is the main focus of this paper.  
In Sec.~\ref{sec:applications} we show this mapping can be used to explore many-body physics, 
and the advantage of using symmetric top molecules versus actual magnetic dipoles.  
In Sec.~\ref{sec:Hfs} we discuss the influence of hyperfine structure on our results, focusing on the case of {\MF} as a typical symmetric top molecule.  
{Finally, in Sec.~\ref{sec:concl}, we conclude.}  
There are three appendices.  Appendix A contains some useful matrix elements for symmetric top molecules.  
Appendix B discusses the relationship between the microscopic angular momentum structure of a highly magnetic atom 
and the model of an elemental magnetic dipole used in the main text.  
Finally, Appendix C discusses the mapping between symmetric top molecules and magnetic dipoles presented in Sec.~\ref{sec:mapping} in second quantization.  
We hope that this makes our results accessible not only to researchers in molecular physics, but also to those in condensed matter.

\section{Rotational structure of symmetric top molecules and coupling to external fields}
\label{sect2}

\subsection{{Quantum mechanics} of symmetric top molecules}
\label{sec:ST}

For quantum gases at ultralow temperatures thermal energy is insufficient to create excitations in the structural or electronic degrees of freedom 
of the atoms or molecules making up the gas.  
Hence, in the theory of ultracold atomic gases, modeling atoms as point-like particles with fixed angular momentum degrees of freedom is appropriate.  
For molecules in low-lying rotational states, we assume that molecules have the equilibrium spatial structures of their nuclei.  
Additionally, we can assume that the molecules remain always in their lowest vibrational and electronic states.  
Hence, the rotational motions of nuclei in closed-shell molecules are well modeled by the quantum mechanics of rigid bodies, 
known as the rigid rotor approximation (RRA)~\cite{Zare_1988}.  
Within the RRA, the rotational energy of a molecule is given by the Hamiltonian 
\beq
\label{eq:Hrot}\hat{H}_{\mathrm{rot}}&=&\frac{\hat{J}_x^2}{2I_x}+\frac{\hat{J}_y^2}{2I_y}+\frac{\hat{J}_z^2}{2I_z}\, ,
\eeq
where $x$, $y$, and $z$ are spatial directions along the principal axes of inertia of the rigid body in the \emph{body-fixed frame}, or \emph{molecule-fixed frame} 
which rotates with the molecule.  We shall also make use of the space-fixed frame of coordinates $X$, $Y$, and $Z$, which is the usual system of laboratory coordinates.  Also in Eq.(~\ref{eq:Hrot}), $\hat{J}_x$, $\hat{J}_y$, and $\hat{J}_z$ are projections of the angular momentum $\hat{\bf{J}}$ 
along the $x$, $y$, and $z$ directions, respectively, and $I_x$, $I_y$, and $I_z$ are the principal moments of inertia.
For a symmetric top molecule, the inertia tensor has a doubly degenerate eigenvalue.  
Defining the $z$-direction as lying along the symmetry axis of the molecule, 
we have that $I_x=I_y=I_{\perp}$. 
Hence, the rotational Hamiltonian for a symmetric top molecule can be written as
\beq
\hat{H}_{\mathrm{rot}}&=&B_0\hat{\mathbf{J}}^2+\left(A_0-B_0\right)\hat{J}_z^2\, ,
\label{H_rot}
\eeq
where we have defined two rotational constants as
\beq
B_0&\equiv&\frac{1}{2I_{\perp}}\, ,\quad A_0~\equiv~\frac{1}{2I_z}\, .
\eeq
One can classify symmetric tops into prolate or oblate tops according to either $A_0>B_0$ or $A_0<B_0$, respectively.  The subscript ``$0$" in the rotational constant sets $B_0$ apart from {the strength of the applied magnetic field $B$}, 
and reminds us that this rotational constant refers to the ground vibrational level of the molecule.  
The square of the total angular momentum $\hat{\mathbf{J}}^2$ is quantized with eigenvalues $J\left(J+1\right)$, 
where the values $J$ are non-negative integers, and the projection along a {molecule-fixed} {quantization axis}, 
which we take to be the $z$-direction, is also quantized with integer eigenvalue $K$.  
For a general symmetric top molecule, the projection $K$ can take on any integer value in the range $-J\le K\le J$.  
{In addition,} the angular momentum is also quantized along the space-fixed $Z$-direction, see Fig.~\ref{fig:Schematic}{(b)}. 
The associated eigenvalue of $\hat{J}_Z$ is denoted by $M$, and takes on integral values {in} $-J\le M\le J$.  
{These} three quantum numbers $J$, $K$, and $M$ completely specify the eigenstates of the symmetric top rigid rotor Hamiltonian Eq.~(\ref{H_rot}).  
The corresponding eigenfunctions are given explicitly in the Euler angle representation by
\beq
\label{eq:STwfcns}
\braket{\omega|JKM}&=&\sqrt{\frac{2J+1}{8\pi^2}}\:\mathscr{D}^{J\ast}_{MK}(\omega)
~,
\eeq
with corresponding eigenenergies, 
\beq
\label{eq:STSpectra}E_{JKM}
&=&
B_0J(J+1)+\left(A_0-B_0\right)K^2
~,
\eeq
where $\mathscr{D}^{J\ast}_{MK}(\omega)$ are the matrix elements of the Wigner $D$-matrix~\cite{Zare_1988} 
that transforms the space-fixed frame onto the molecule-fixed frame by three Euler angles $\omega=(\phi,\theta,\chi)$.  Throughout this paper, the appearance of $\omega$ in any $D$-matrix or spherical harmonic refers to the three Euler angles rotating the molecule-fixed frame to the space-fixed frame and not to an angular frequency.  States with the same $\left|K\right|\neq 0$ are doubly degenerate, due to the cylindrical symmetry of the Hamiltonian Eq.~(\ref{H_rot}).  
For a discussion of the degree to which the $\pm |K|$ degeneracy persists for symmetric top molecules beyond the RRA, see {Sec.~\ref{sec:Hfs}}.

\begin{figure}[tbp]
\centerline{\includegraphics[width=0.6\columnwidth]{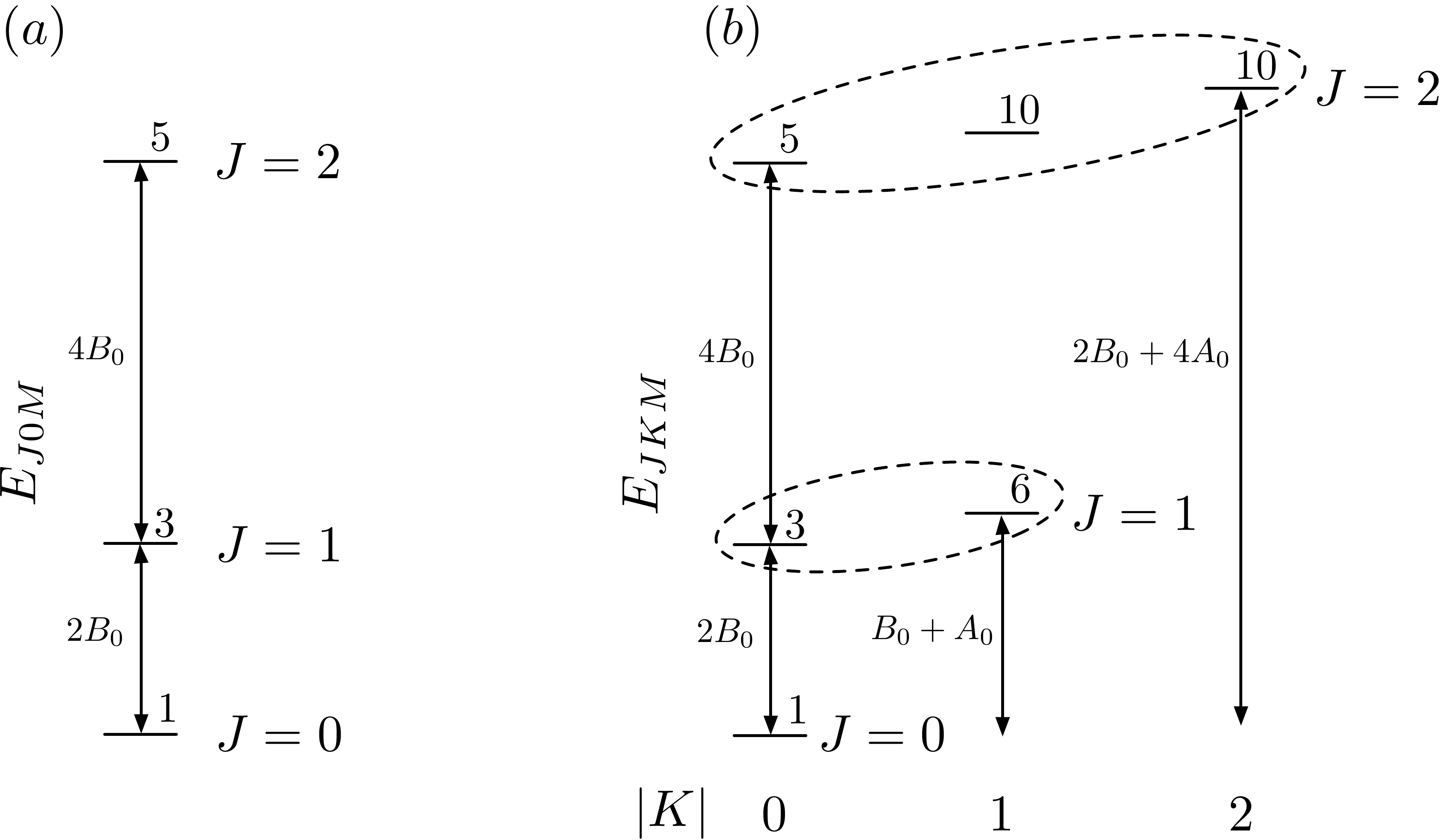}}
\caption{\label{fig:Energies}  
{Rotational energy spectra} of (a) linear rigid rotors $E_{J0M}$ and (b) symmetric top molecules $E_{JKM}$.  
The degeneracy of a given level is denoted by the number to the upper right of the level.  
The dashed ellipses in (b) are guides to the eye of fixing $J$ and varying $\left|K\right|$, called fixed $J$-manifolds.  
For simplicity, panel (b) assumes a prolate top with $A_0>B_0$.  
However, the scale of $A_0/B_0$ for a given symmetric top molecule may be very different than the one displayed.}
\end{figure}

For the case of a linear rigid rotor, the rotational angular momentum must be perpendicular to the symmetry axis of the molecule, 
and {thus} its projection $K$ along the symmetry axis is identically zero, as shown Fig.~\ref{fig:Schematic}{(a)}.  
In this case, the wave functions become proportional to spherical harmonics $Y_{JM}(\theta,\phi)$
\beq
\braket{\omega|J0M}&=&\frac{1}{\sqrt{2\pi}}Y_{JM}(\theta,\phi)
\label{eq:J0M}
~,
\eeq
with corresponding eigenenergies, 
\beq
\label{eq:LMSpectra} 
E_{J0M}&=&B_0J(J+1)
~.
\eeq
The disappearance of the angle $\chi$ from the wavefunction Eq.~(\ref{eq:J0M}) is a consequence of the linear structure.  The two spectra Eq.~(\ref{eq:STSpectra}) and Eq.~(\ref{eq:LMSpectra}) and their associated degeneracies are compared in Fig.~\ref{fig:Energies}.

\subsection{The Stark effect and matrix elements of the dipole moment}
\label{sec:Stark}
\begin{figure}[tbp]
\centerline{\includegraphics[width=0.9\columnwidth]{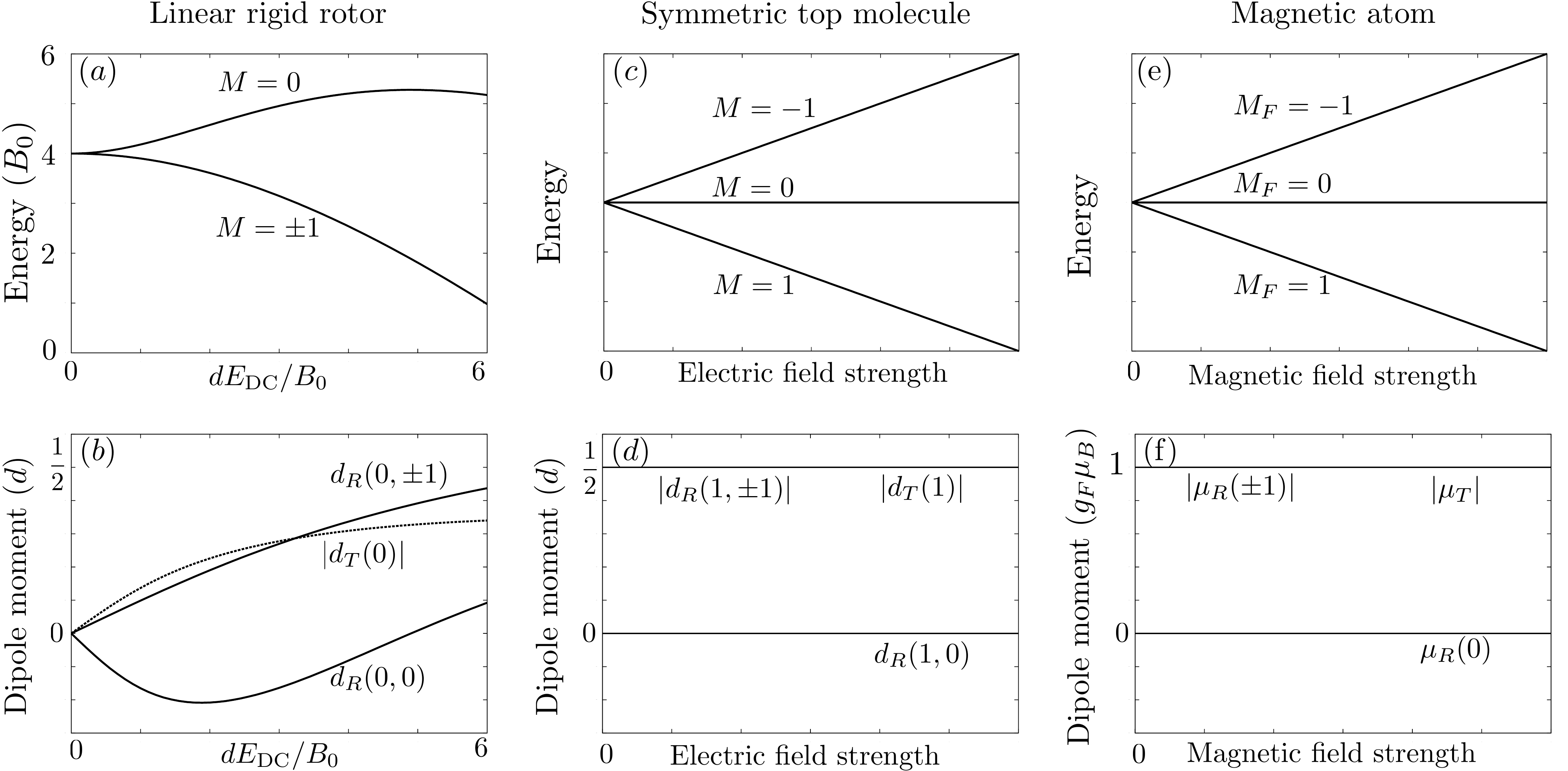}}
\caption{\label{fig:singsigvsST}  
Energies and dipole matrix elements of a linear rigid rotor in the states $|\bar{1}0M\rangle$ (panels (a)-(b)), a symmetric top molecule in the states $|\bar{1}1M\rangle$ (panels (c)-(d)), and a magnetic atom in the states $|F=1,M_F\rangle$ (panels (e)-(f)).  The behavior of the energies and electric dipole matrix elements of a symmetric top molecule in an electric field match the behavior of the energies and magnetic dipole matrix elements of a magnetic dipole in a magnetic field.  The behavior of a linear rigid rotor in an electric field is significantly different.  The dipole matrix elements are given in terms of the transition matrix elements $d_T(K)\equiv \langle \bar{1}K,\pm 1|\hat{d}_{\pm 1}|\bar{1}K0\rangle$ ($\mu_T\equiv \langle 1,\pm 1|\hat{\mu}_{\pm 1}|10\rangle$ for the magnetic dipole) and the resonant matrix elements $d_R(K,M)\equiv \langle \bar{1}KM|\hat{d}_0|\bar{1}KM\rangle$ ($\mu_R(M_F)\equiv \langle 1M_F|\hat{\mu}_0|1M_F\rangle$ for the magnetic dipole).  See Sec.~\ref{sec:magdipoles} for a complete discussion of magnetic dipole matrix elements.
}
\end{figure}

Applying a DC electric field to polar molecules causes the Stark effect, which is described perturbatively via an interaction term between 
the dipole moment operator of molecules $\hat{\bf d}$ and applied DC electric field ${\bf E}_{\rm DC}$, that is,
\beq
\hat{H}_{\rm Stark}&=&-\hat{\bf d}\cdot{\bf E}_{\rm DC} 
~.
\eeq
In the following we assume that the DC electric field is applied along the space-fixed $Z$-direction {$\mathbf{e}_Z$}
with a strength of $E_{\mathrm{DC}}$, i.~e., $\mathbf{E}_{\mathrm{DC}}=E_{\mathrm{DC}}\mathbf{e}_Z$.  
For symmetric top molecules in the RRA, first-order perturbation theory yields the energy shift as 
\beq
\Delta E^{(1)}_{JKM}&=&\bra{JKM}\hat{H}_{\rm Stark} \ket{JKM}
\nonumber\\
&=&-d E_{\mathrm{DC}} \frac{KM}{J(J+1)}
\label{eq:FOSE}
~, 
\eeq
where $d$ is the permanent electric dipole moment of the molecule{, see Appendix A}.  
For a fixed $J$, the $(2J+1)$-fold degeneracy indexed by $M$ has been lifted by the Stark effect, but there still remains a $2$-fold degeneracy in the product of $K$ and $M$ when $|K|>0$.  
For a symmetric top molecule there is a linear Stark effect whenever $KM\neq0$.  
That is, the Stark effect is linear both in $E_{\mathrm{DC}}$ and $M$ for states with the same $J$ and $K$, i.e., a fixed $(J,K)$-manifold.  
We will call states with $KM\neq0$ \emph{precessing states}, 
due to the fact that classically a state with $KM\ne 0$ precesses rather than tumbles~\cite{Precess,Lemeshko_Friedrich}.  
This precession is what gives a nonzero expectation of the dipole moment along the field direction and, hence, the first-order Stark effect.  
In addition, as the matrix elements Eq.~(\ref{eq:FOSE}) are the only ones which are non-vanishing for a given $J$ and $\left|K\right|$, 
the states $|JKM\rangle$ are the eigenstates of the rotation and Stark Hamiltonian up to first order in $dE_{\mathrm{DC}}/B_0$.  
For discussion of dipole matrix elements, it is convenient to define the space-fixed spherical basis $\{\mathbf{e}_p\}_{p=0,\pm1}$, 
given by the unit vectors~\cite{Zare_1988}
\beq
\mathbf{e}_{\pm1}&\equiv&\mp\left(\mathbf{e}_X\pm i\mathbf{e}_Y\right)/\sqrt{2}\, ,\quad
\mathbf{e}_0~\equiv~ \mathbf{e}_Z\, .
\label{eq:sphbasis}
\eeq
The expectation of the dipole operator along space-fixed spherical direction $\mathbf{e}_p$, 
$\hat{d}_p\equiv \hat{\mathbf{d}}\cdot \mathbf{e}_p$, in the basis $|JKM\rangle$ with $J$ fixed, is
\beq
\langle JK'M'|\hat{d}_p|JKM\rangle&=&d\left(-1\right)^{J-M'}\tj{J}{-M'}{1}{p}{J}{M}K\sqrt{\frac{2J+1}{J\left(J+1\right)}}\:\delta_{KK'}\, ,
\label{eq:dipmatelems}
\eeq
where$(\,{}_{\dots}^{\dots}\,)$ denotes the Wigner  3-$j$ symbol~\cite{Zare_1988}.  The fact that all dipole matrix elements are diagonal in $K$ means that any two states with projections $(K,M)$ and $(-K,-M)$ do not mix under electric field or dipole-dipole interactions even though they are degenerate in the rotational and Stark energy spectrum.  Hyperfine interactions which break this degeneracy are discussed in Sec.~\ref{sec:Hfs}.  The dipole matrix elements Eq.~(\ref{eq:dipmatelems}) can be written in the form of the Wigner-Eckart theorem for a spherical tensor operator
\beq
\langle JK'M'|\hat{d}_p|JKM\rangle&=&\left(-1\right)^{J-M'}\tj{J}{-M'}{1}{p}{J}{M}\langle JK'|\!|\hat{\mathbf{d}}|\!|JK\rangle\, ,
\label{eq:WEform} 
\eeq
via identification of the reduced matrix element
\beq
\langle JK'|\!|\hat{\mathbf{d}}|\!|JK\rangle&=&dK\sqrt{\frac{2J+1}{J\left(J+1\right)}}\:\delta_{KK'}\, .
\label{eq:redmatelem} 
\eeq
There are three important implications of the linear Stark effect for our purposes.  
First, according to Eq.~(\ref{eq:FOSE}), the energetic separation between states $|JKM\pm1\rangle$ and $|JKM\rangle$ has the same magnitude for all values of $M$.  
Second, for any non-vanishing electric field strength $E_{\mathrm{DC}}$, 
we obtain an expected dipole moment along the space-fixed $Z$-axis of $dKM/[J(J+1)]$, which becomes half the permanent dipole moment when $J=KM=1$.  
Finally, due to the spherical tensor structure in Eq.~(\ref{eq:WEform}), the dipole operator behaves as though it simply points in the direction of $\mathbf{J}$ 
when our attention is restricted to a single $(J,K)$-manifold.  
This mimics the behavior of magnetic dipoles, as will be discussed in Sec.~\ref{sec:magdipoles}.  
These features of symmetric top molecules are compared with the behavior of a linear rigid rotor in Fig.~\ref{fig:singsigvsST}.

We now contrast these results for symmetric top molecules with the results for a linear rigid rotor.  
For a linear rigid rotor the first-order perturbation Eq.~(\ref{eq:FOSE}) identically vanishes for any rotational eigenstate since $K=0$.  
Then, the second-order Stark shift is given as
\beq
\Delta E^{(2)}_{JKM}&=&
\sum_{(J'\!,K'\!,M')\neq (J,K,M)}\frac{|\bra{JKM}\hat{H}_{\rm Stark} \ket{J'K'M'}|^2}{E_{JKM}-E_{J'K'M'}}
\\
&=&
\frac{(d E_{\mathrm{DC}})^2}{2B_0}\left\{
-\frac{\left[(J+1)^2-K^2\right]\left[(J+1)^2-M^2\right]}{(J+1)^3(2J+1)(2J+3)}
+\frac{\left(J^2-K^2\right)\left(J^2-M^2\right)}{J^3(2J+1)(2J-1)}
\right\}
~, \nonumber
\eeq
which yields for $K=0$, 
\beq
\Delta E^{(2)}_{J0M}&=&\frac{(d E_{\mathrm{DC}})^2}{2B_0}\left\{\frac{J(J+1)-3M^2}{J(J+1)(2J-1)(2J+3)}\right\}~. 
\eeq
Here, pairs of states with the same $\left|M\right|$ remain degenerate.  
The expected dipole moment of a linear rigid rotor along the $Z$-direction can be obtained 
perturbatively from the Feynman-Hellman theorem~\cite{lebellac2006} as
\beq
\langle J0M|\hat{d}_{{p=0}}|J0M\rangle&=&\frac{d^2E_{\mathrm{DC}}}{B_0}\left\{\frac{J(J+1)-3M^2}{J(J+1)(2J-1)(2J+3)}\right\}\, .
\eeq
As opposed to the space-fixed dipole moments of a symmetric top Eq.~(\ref{eq:dipmatelems}), which are independent of $E_{\mathrm{DC}}$ for small fields, 
the dipole moments of a linear rigid rotor depend linearly on $E_{\mathrm{DC}}$, and so are small in fields which are small compared to $B_0/d$.  
On the other hand, for fields which are large compared to $B_0/d$ the expected dipole moments approach the permanent value $d$, 
while states with different values of $\left|M\right|$ are separated by energies of the order of the rotational constant $B_0$.  Because a linear rigid rotor requires non-perturbative fields in order to access a significant fraction of the permanent dipole moment along the field direction, 
the eigenstates of the rotational and Stark Hamiltonian for rigid rotors are no longer well-approximated by the rotational eigenstates $|J0M\rangle$.  
However, due to the facts that there are no crossings between states which correspond to different rotational quanta $J$ in zero field 
and that $M$ is a good quantum number even in the presence of $\mathbf{E}_{\mathrm{DC}}$, 
we can label the states as $|\bar{J}0M\rangle$, where $\bar{J}$ is an adiabatic label correlating to zero field.
\begin{table}
\caption{\label{table:SingSig} 
The  exchange statistics (f=fermion, b=boson), rotational constants $B_0$, permanent dipole moments $d$, and polarizing fields $E_p$ such that $dE_p/B_0=6$ 
for several relevant $^1\Sigma$ alkali metal dimers~\cite{Aldegunde,Hfs,MHH}.  
At the field $E_p$, the expected dipole moment of the ground state along the field direction obtains roughly $70\%$ of its permanent value.  The large fields $E_p$ required to polarize the dipole moment are in contrast to symmetric tops, in which essentially no field is required to polarize a large fraction of the dipole moment.
}
\begin{center}
\begin{tabular}{|c|c|c|c|c|}
\hline &Statistics&$B_0$ (GHz) & $d$ (Deybe) & $E_p$ (kV/cm)\\
\hline \LiCs&f& 6.520&5.52&14.07\\
\hline \NaK&f&2.826 &2.76&12.20\\
\hline \RbCs&b&0.504&1.25&4.80\\
\hline \KRb&f&1.114&0.566&23.4\\
\hline \LiNa&f&12.735&0.56&271.043\\
\hline
\end{tabular}
\end{center}
\end{table}
\begin{table}
\caption{\label{table:SymmTop} 
The  exchange statistics, rotational constants $B_0$, permanent dipole moments $d$, and critical fields $E_c$ such that $dE_c/B_0=1$ 
for selected symmetric top molecules~\cite{NIST,Townes_Schawlow,Wei_Kais_11}.  
The critical field is an estimate for significant breakdown of the linear Stark effect regime for $J=1$ states.  Thus, a tremendous range of electric fields can be used in experiments without going beyond the linear Stark approximation.
}
\begin{center}
\begin{tabular}{|c|c|c|c|c|c|}
\hline &Statistics&$B_0$ (GHz) & $d$ (Deybe) &$E_c$ (kV/cm)\\
\hline $^{12}${\MF}&b& 25.536&1.850&27.419\\
\hline  $^{13}${\MF}&f& 24.862&1.850&26.696\\
\hline CH$_3$Cl&b& 13.292&1.87&14.120\\
\hline CH$_3$I&b& 7.501&1.62&9.19\\
\hline CH$_{3}$CN&f&9.1988&3.92&4.661\\
\hline
\end{tabular}
\end{center}
\end{table}

The behavior of linear rigid rotors and symmetric top molecules in a static electric field are contrasted in Tables~\ref{table:SingSig} and \ref{table:SymmTop} and Fig.~\ref{fig:singsigvsST}.  The parameters in Table~\ref{table:SingSig} are obtained from experiment for KRb, but are computed from density functional theory or other ab initio means for the other species.  In contrast, the parameters in Table~\ref{table:SymmTop} are obtained from experiment due to the ready availability of these symmetric top species.  As shown in Table~\ref{table:SingSig} the fields required to significantly orient the ground state of alkali dimer molecules along the field direction 
via the second-order Stark effect range from a few to a few hundred kV/cm.  
In contrast, due to the linear Stark shift, the symmetric top molecules given in Table~\ref{table:SymmTop} obtain 
half of their permanent dipole moment for the $|11-1\rangle$ state in any non-vanishing field.  
The critical fields $E_c$ given in Table~\ref{table:SymmTop} represent a rough estimate where the purely linear Stark effect, i.~e., 
first order perturbation theory, becomes of the same order of magnitude as the second-order Stark effect for $J=1$.  Note that the critical field $E_c$ where breakdown of first-order perturbation theory occurs depends on the particular state $|JKM\rangle$.  A comparison between the energetic spectrum and dipole moments of a symmetric top with $J=K=1$, a linear rigid rotor with $J=1$, and a magnetic atom with $F=1$ are given 
in Fig.~\ref{fig:singsigvsST}.  
  While panels (a) and (b) of Fig.~\ref{fig:singsigvsST} are numerical data from non-perturbative calculations, panels (c)-(f) are schematics of the behavior where linear coupling to the applied field holds, see Tables~\ref{table:SingSig} and \ref{table:SymmTop}.  Figure \ref{fig:singsigvsST} stresses the similarity between symmetric top molecules and magnetic dipoles both in terms of energies and dipole matrix elements, and also stresses the difference between linear rigid rotors and magnetic dipoles.  In the strong-field regime where first-order perturbation theory breaks down, symmetric tops will also display second and higher-order Stark effects.  There, the symmetric top spectra will appear more like those of linear rigid rotors, eventually becoming high-field seeking pendular states.  A discussion of symmetric tops in strong electric fields may be found in Ref.~\cite{Haertelt_Friedrich_08}.

\subsection{Interaction of symmetric top molecules with other external fields}
\label{sec:otherfields}
In this section we discuss two other means by which symmetric top molecules couple to external fields.  
The first is the coupling of symmetric top molecules to off-resonant optical potentials.  
Optical trapping is important for our proposal, as it is the only means 
by which we can trap both the high-field and low-field seeking states contained in a particular $(J,K)$-manifold simultaneously.  The second topic is the coupling of magnetic fields to the magnetic moment generated by rotation of the molecule, known as the rotational Zeeman effect.   Magnetic fields are an important means of separating out hyperfine structure such that single quantum states may be addressed~\cite{Hfs}, and so it is also important to understand the effects of magnetic fields on the rotational structure of the molecule.  The Zeeman effect due to intrinsic nuclear magnetic moments and associated nuclear hyperfine structure will be discussed in Sec.~\ref{sec:Hfs}.

Let us consider the application of an optical field
\beq
\mathbf{E}_{\mathrm{opt}}\left(\mathbf{r},t\right)&=&
\mathbf{E}_{\mathrm{opt}}\left(\mathbf{r}\right)e^{-i\omega_{\mathrm{opt}}t}
+\mathbf{E}_{\mathrm{opt}}^{\star}\left(\mathbf{r}\right)e^{i\omega_{\mathrm{opt}}t}\, ,
\eeq
where the frequency of the field $\omega_{\mathrm{opt}}$ is far detuned from any molecular resonances.  
We will suppose that the field separates into spatial and polarization components as
\beq
\label{eq:fielddef}\mathbf{E}_{\mathrm{opt}}\left(\mathbf{r}\right)&=&E_{\mathrm{opt}}\left(\mathbf{r}\right)\sum_{p=-1}^{1}\varepsilon_{p}\mathbf{e}_p^{\star}\, ,
\eeq
where $\mathbf{e}_p$ are the spherical basis vectors Eq.~(\ref{eq:sphbasis}) 
and the polarization vector $\boldsymbol{\varepsilon}$ has unit norm, $\boldsymbol{\varepsilon}^{\star}\cdot\boldsymbol{\varepsilon}=1$.  
When the field is far off-resonant, the coupling of the molecule to the field may be described as
\beq
\label{eq:HoptStark} \hat{H}_{\mathrm{opt}}&=&
-\mathbf{E}^{\star}_{\mathrm{opt}}\left(\mathbf{r}\right)\cdot \tilde{\alpha}\left(\omega_{\mathrm{opt}}\right)\cdot \mathbf{E}_{\mathrm{opt}}\left(\mathbf{r}\right)\, ,
\eeq
where $\tilde{\alpha}\left(\omega_{\mathrm{opt}}\right)$ is the dynamical polarizability tensor~\cite{Bonin_Kresin} 
evaluated at the field frequency $\omega_{\mathrm{opt}}$.  The symmetry of the polarizability tensor is the same as the inertia tensor.  
That is, $\tilde{\alpha}\left(\omega_{\mathrm{opt}}\right)$ also has a doubly degenerate eigenvalue, and the principal axes of the two tensors are parallel.  
It should be noted, however, that the polarizability tensor need not be prolate even if the inertia tensor is; only the symmetry is specified \emph{a priori}.  
The cylindrical symmetry implies that the polarizability tensor may be written in the molecule-fixed spherical basis $\{\mathbf{e}_q\}_{q=0,\pm1}$ as
\beq
\tilde{\alpha}\left(\omega_{\mathrm{opt}}\right)&=&
\alpha_{\parallel}\left(\omega_{\mathrm{opt}}\right)\mathbf{e}_{q=0}\otimes \mathbf{e}_{q=0}
+\alpha_{\perp}\left(\omega_{\mathrm{opt}}\right)\left[\mathbf{e}_{q=1}\otimes \mathbf{e}_{q=1}^{\star}+\mathbf{e}_{q=-1}\otimes \mathbf{e}_{q=-1}^{\star}\right]\, ,
\eeq
where the molecule-fixed $z$-axis is defined to be along the symmetry axis.  
Transforming from the molecule-fixed frame indexed by $q$ to the space-fixed frame indexed by $p$ via the transformation~\cite{Brown_Carrington}
\beq
\mathbf{e}_{q}&=&\sum_{q=-1}^1\mathscr{D}^{1}_{pq}\left(\omega\right)\mathbf{e}_p\, ,
\eeq
with $\omega$ the Euler angles connecting the two frames, the Hamiltonian Eq.~(\ref{eq:HoptStark}) may be written
\beq
\hat{H}_{\mathrm{opt}}&=&
-\left|E_{\mathrm{opt}}\left(\mathbf{r}\right)\right|^2
\left[
\bar{\alpha}\left(\omega_{\mathrm{opt}}\right)+\Delta \alpha\left(\omega_{\mathrm{opt}}\right)\sum_{p=-2}^{2}C^{\left(2\right)}_{p}(\omega)\Upsilon_{p}
\right]\, .
\label{eq:Hoptelems} 
\eeq
Here we have defined products of polarization components $\Upsilon_{p}$ as
\beq
\Upsilon_{\pm 2}&=&-\sqrt{\frac{2}{3}}\varepsilon_{\mp 1}\varepsilon_{\pm 1}^{\star}\, ,\quad
\Upsilon_{\pm 1}=\frac{1}{\sqrt{3}}\left(\varepsilon_0\varepsilon_{\pm1}^{\star}-\varepsilon_{\mp1}\varepsilon_0^{\star}\right)\, ,\quad 
\Upsilon_0=\left|\varepsilon_0\right|^2-1/3\, ,
\label{Upsilons}
\eeq
and unnormalized spherical harmonics $C^{\left(\ell\right)}_p$ as
\beq
\label{eq:unnSH}C^{\left(\ell\right)}_p\left(\omega\right)&=&\sqrt{\frac{4\pi}{2\ell+1}}Y_{\ell p}\left(\theta,\phi\right)\, ,
\eeq
Also, we introduced the polarizability tensor invariants
\beq
\label{eq:poldef}\bar{\alpha}\left(\omega_{\mathrm{opt}}\right)&\equiv& \frac{2\alpha_{\perp}\left(\omega_{\mathrm{opt}}\right)+\alpha_{\parallel}\left(\omega_{\mathrm{opt}}\right)}{3}\, ,
\quad \Delta\alpha\left(\omega_{\mathrm{opt}}\right)~\equiv~ \alpha_{\parallel}\left(\omega_{\mathrm{opt}}\right)-\alpha_{\perp}\left(\omega_{\mathrm{opt}}\right)\, .
\eeq
Note that all of the components of the polarization vector $\boldsymbol{\varepsilon}$ in Eq.~(\ref{Upsilons}) are given in the space-fixed frame. 
For {\MF}, the static values of the polarizability tensor have been determined experimentally to be 
$\bar{\alpha}\left(0\right)=2.540\:\mathrm{\AA}^3$~\cite{Polarizability} 
and $\Delta\alpha\left(0\right)/\bar{\alpha}\left(0\right)=0.33$~\cite{Haertelt_Friedrich_08}.  The average polarizability $\bar{\alpha}$ is smaller 
than the $\sim 150\: \mathrm{\AA}^3$ values typical of the alkali metal dimers~\cite{Deiglmayr_Aymar_08,Kotochigova_Demille_10}.  As the electronic structure of {\MF} lies in the deep ultraviolet~\cite{Locht_Leyh}, 
optical trapping at the common frequencies of 1064nm or 532nm appears to be reasonable.

For fields whose polarization is parallel to a spherical unit vector, the factors $\Upsilon_{\pm2}$ and $\Upsilon_{\pm1}$ all vanish identically, 
and so $M$ remains a good quantum number.  However, even in this case states with different $|M|$ experience different depths of the optical potential 
due to the tensor shift depending on $C^{\left(2\right)}_0\left(\omega\right)$.  
As the typical depth of an optical potential ($\le$ 10-100 kHz) is much less than the splittings between rotational levels($\sim$GHz), 
we can consider only the diagonal terms of Eq.~(\ref{eq:Hoptelems}), where the expected tensor shift is proportional to
\beq
\langle JKM|C^{\left(2\right)}_0\left(\omega\right)|JKM\rangle&=&
\frac{\left[J\left(J+1\right)-{3K^2}\right]\left[J\left(J+1\right)-{3M^2}\right]}{J\left(J+1\right)\left(2J-1\right)\left(2J+3\right)}\,.
\label{eq:tensshift}
\eeq
We expect that many of the techniques which have been devised to cancel the tensor shift in linear rigid rotors~\cite{Gorshkov_Manmana_11b} 
can also be adapted to the case of symmetric top molecules.  For example, one can choose the polarization such that $\Upsilon_0=0$~\cite{Kotochigova_Demille_10}, or use multiple optical fields with different tensor shifts to either cancel the shift altogether or make it spatially independent~\cite{Gorshkov_Manmana_11b}.  Other techniques which alter the character of the internal states, such as microwave dressing, can also cancel the tensor shift.  Techniques which alter the internal state character will generally destroy the spherical tensor nature of the dipole matrix elements which is key to our correspondence.  However, it is still possible to cancel the tensor shift while keeping our correspondence by exploiting the geometry of the polarizations or using multiple optical fields. 

We now turn to the rotational Zeeman effect, which is described by the Hamiltonian
\beq
\hat{H}_{\mathrm{B;rot}}&=&-\mu_N\mathbf{B}\cdot\tilde{g}\cdot\hat{\mathbf{J}}\, ,
\label{eq:rotZeeman}
\eeq
where $\mu_N$ is the nuclear magneton.  The tensor $\tilde{g}$ describes the anisotropy of the coupling between rotation and the external magnetic field.  
As was also the case for the dynamical polarizability above, the tensor $\tilde{g}$ has the same symmetry as the inertia tensor, 
and may be fully described by two elements $g_{\parallel}$ and $g_{\perp}$.  
The parameters for {\MF}, $g_{\parallel}=0.265$ and $g_{\perp}=-0.062$, have been determined experimentally~\cite{ZeemanParams}.  
Taking a similar approach as for the AC Stark shift above, we find that the Hamiltonian Eq.~(\ref{eq:rotZeeman}) may be written as
\beq
\hat{H}_{\mathrm{B;rot}}&=&-\mu_N\left[\bar{g}\mathbf{B}\cdot\hat{\mathbf{J}}+\Delta g\sum_{p=-2}^{2}C^{\left(2\right)}_p\left(\omega\right)\gamma_p\right]\, ,
\label{eq:HrotZ}
\eeq
where
\beq
\gamma_{\pm2}&=&\sqrt{\frac{2}{3}}B_{\mp1}\hat{J}_{\mp1}\, ,\quad 
\gamma_{\pm1}~=~-\frac{1}{\sqrt{3}}\left(B_0\hat{J}_{\mp 1}+B_{\mp 1}\hat{J}_0\right)\, ,\;\; 
\label{gammas1}\\
\gamma_0&=&\frac{1}{3}\left(2B_0\hat{J}_0+B_{1}\hat{J}_{-1}+B_{-1}\hat{J}_1\right)\, ,
\label{gammas2}
\eeq
are linear combinations of rotation operators in the space-fixed frame and $B_p\equiv \mathbf{B}\cdot \mathbf{e}_p$ are the components of the magnetic field also in the space-fixed frame.  Also,  we have defined the tensor invariants
\beq
\bar{g}\equiv \frac{2g_{\perp}+g_{\parallel}}{3}\, ,\;\;\Delta g\equiv g_{\parallel}-g_{\perp}\, .
\eeq
As the nuclear magneton $\mu_N\approx 762$ Hz/Gauss, 
the rotational Zeeman effect will not appreciably mix rotational levels for any reasonable value of the magnetic field, 
and so we can consider only the matrix elements diagonal in $J$.  
Assuming that the field is along the $\mathbf{e}_Z$ direction, $\mathbf{B}=B\mathbf{e}_Z$, we find the matrix elements
\beq
\langle JK'M'|\hat{H}_{\mathrm{B;rot}}|JKM\rangle&=&-\mu_NB M\left[g_{\perp}+\Delta g\frac{K^2}{J\left(J+1\right)}\right]\delta_{MM'}\delta_{KK'}\, .
\label{eq:RotZeeman}
\eeq
While the matrix elements for the linear Stark effect, Eq.~(\ref{eq:FOSE}), depend only on the product $KM$, the matrix elements of the rotational Zeeman interaction depend on $K$ and $M$ separately, and so sorting of the state specified by $J$, $K$, and $M$ is possible via a combination of electric and magnetic fields.

\section{Relationship between symmetric top molecules and magnetic dipoles}
\label{sec:Reln}

In this section, we first review the single-particle physics of a magnetic atom by showing 
the explicit forms of magnetic dipole moments and the Zeeman effect. 
Then, we show that there is a rigorous correspondence between a magnetic atom and a symmetric top molecule which we present as the mapping in Table~\ref{table:mapping}.
Finally, we extend our correspondence to include the dipole-dipole interaction, and show that dipole-dipole interactions are significantly larger for the simulated magnetic dipole than for naturally occurring magnetic dipoles.  For example, the symmetric top molecule ${\rm CH_3F}$ can simulate the magnetic atom ${}^{52}{\rm Cr}$ 
with a factor of 620 enhancement in the dipole-dipole interaction energy. 
The correspondence shown in this section can be applied to a large class of quantum magnets, such as magnetic atoms, long-range spin models, and molecular magnets, as will be discussed in Sec.~\ref{sec:applications}. 

\subsection{Magnetic dipoles and the Zeeman effect}
\label{sec:magdipoles}

Let us consider, as a model of a magnetic atom, e.g., ${}^{52}{\rm Cr}$, ${}^{168}{\rm Er}$, or ${}^{164}{\rm Dy}$, 
a point particle with an angular momentum $\mathbf{F}$ and a magnetic dipole operator $\hat{\boldsymbol{\mu}}$ 
which is proportional to this angular momentum
\beq
\label{eq:mueff}\hat{\boldsymbol{\mu}}&=&g_F\mu_B\hat{\mathbf{F}}\, ,
\eeq
where $g_F$ is the effective $g$-factor.  
The appropriate basis for describing this object is $|FM_F\rangle$, where $F\left(F+1\right)$ is the eigenvalue of $\hat{\mathbf{F}}^2$ 
and $M_F$ is the eigenvalue of the projection of $\hat{\mathbf{F}}$ on the space-fixed $Z$-axis.  
The details of the connection between the effective description of the magnetic dipole, Eq.~(\ref{eq:mueff}), 
and the microscopic angular momentum structure of a highly magnetic atom are presented in {Appendix B}.  
The magnetic dipole moment couples to an external magnetic field $\mathbf{B}$ via the Zeeman Hamiltonian, 
\beq
\hat{H}_{\mathrm{Zeeman}}&=&-\hat{\boldsymbol{\mu}}\cdot\mathbf{B}\, .
\eeq
The matrix elements of the magnetic dipole operator can be evaluated using the Wigner-Eckart theorem~\cite{Zare_1988} as
\beq
\label{eq:magdips}
\langle FM'_{F}|\hat{\mu}_p|FM_F\rangle
&=&
 \left(-1\right)^{F-M_{F}'}\tj{F}{-M'_{F}}{1}{p}{F}{M_F}
g_F\mu_B \langle F|\!|\hat{\bf F}|\!|F\rangle 
\, ,
\eeq
where the reduced matrix element is
\beq
\langle F|\!|\hat{\bf{F}}|\!|F\rangle 
&=&
\sqrt{F\left(F+1\right)\left(2F+1\right)}
\, .
\label{eq:magredmat}
\eeq
In particular, this gives the first-order Zeeman shifts for a magnetic field along 
the space-fixed $Z$-direction $\boldsymbol{e}_Z$, i.e., 
${\boldsymbol B}=B\boldsymbol{e}_Z$ as
\beq
\bra{FM_F}\hat{H}_{\rm Zeeman} \ket{FM_F}&=&-g_FM_F  \mu_B   B
\label{eq:FOZE}
~. 
\eeq 
Hence, insofar as the Zeeman effect is not of the same order of magnitude as the distance between adjacent $F$-manifolds, 
we find that the eigenstates of a magnetic dipolar particle in a magnetic field are $|FM_F\rangle$ with energies given by Eq.~(\ref{eq:FOZE}) 
and dipole matrix elements given by Eq.~(\ref{eq:magdips}).

\subsection{Mapping between magnetic dipoles and symmetric top molecules}
\label{sec:mapping}

We are now in a position to demonstrate a mapping between symmetric top molecules and magnetic atoms 
which allows us to perform quantum simulation of a gas of the latter with the former.  
We begin by collecting the main results from Secs.~\ref{sect2} and \ref{sec:magdipoles}.  
To first order in $dE_{\mathrm{DC}}/B_0$, the eigenstates of a symmetric top molecule in an electric field $\mathbf{E}_{\mathrm{DC}}=E_{\mathrm{DC}}\mathbf{e}_Z$ 
are the states $|JKM\rangle$ with energies and dipole moments given by
\beq
\left(\hat{H}_{\mathrm{rot}}+\hat{H}_{\mathrm{Stark}}\right)|JKM\rangle&=&
\left[B_0J\left(J+1\right)+\left(A_0-B_0\right)K^2-\frac{dK}{J\left(J+1\right)}M E_{\mathrm{DC}}\right]|JKM\rangle
\, ,\qquad\label{eq:totalEE}\\
\label{eq:totalED}\langle JKM'|\hat{d}_p|JKM\rangle&=&\left(-1\right)^{J-M'}\tj{J}{-M'}{1}{p}{J}{M} dK\sqrt{\frac{2J+1}{J\left(J+1\right)}}\, .
\eeq
For a fixed $(J,K)$-manifold, the rotational energy $B_0J\left(J+1\right)+\left(A_0-B_0\right)K^2$ is common to all states and so may be taken as the zero of energy.  For a magnetic dipole with angular momentum ${F}$ in a magnetic field $\mathbf{B}=B\mathbf{e}_Z$, 
the eigenstates are $|FM_F\rangle$ with energies and magnetic dipole moments given by
\beq
\label{eq:totalME}\hat{H}_{\mathrm{Zeeman}}|FM_F\rangle&=&-g_F\mu_B M_F B|FM_F\rangle \, ,\\
\label{eq:totalMD}\langle FM'_{F}|\hat{\mu}_p|FM_F\rangle&=& \left(-1\right)^{F-M_{F}'}\tj{F}{-M'_{F}}{1}{p}{F}{M_F}g_F\mu_B \sqrt{F\left(F+1\right)\left(2F+1\right)}\, .
\eeq
Comparing the electric dipole matrix elements Eq.~(\ref{eq:totalED}) with the magnetic dipole matrix elements Eq.~(\ref{eq:totalMD}), 
we see that the symmetric top molecule has an electric dipole moment 
which behaves exactly as a magnetic dipole moment provided that we make the identifications 
\beq
\label{eq:mapping1}
J&\rightarrow&F\, ,\qquad 
M~\rightarrow~M_F \, , \qquad 
\frac{dK}{F\left(F+1\right)}~\rightarrow~g_F\mu_B\, .
\eeq
We use the notation ``$\rightarrow$" to denote the mapping, as the two sides clearly do not have matching units.  
What we mean in the mapping is that the two systems obey the same algebra provided that we identify the coefficients according to Eq.~(\ref{eq:mapping1}).  
Now, comparing the energy of the symmetric top molecule Eq.~(\ref{eq:totalEE}) in a fixed $(J,K)$-manifold and the energy of the magnetic dipole Eq.~(\ref{eq:totalME}) 
while applying the mapping Eq.~(\ref{eq:mapping1}), we find that the energies are the same provided we make the identification 
\beq
\label{eq:mapping2}
 E_{\mathrm{DC}}&\rightarrow& B\, .
\eeq
Thus, a symmetric top molecule in a fixed $(J,K)$-manifold and a magnetic dipole have the same single-particle energies and dipole moments, 
when the dipole moments and fields are mapped according to Eqs.~(\ref{eq:mapping1}) and (\ref{eq:mapping2}).

To complete the mapping, we will define the relationship between the interaction of two symmetric top molecules in the same $(J,K)$-manifold 
and the interaction of two magnetic dipoles. 
For this purpose, it is useful to recast the dipole-dipole interaction Eq.~(\ref{eq:HDDI1}) as a contraction of two rank-two spherical tensors as~\cite{Brown_Carrington}
\beq
\label{eq:HDDIv2}\hat{H}_{\mathrm{DDI};\mathcal{D}}&=&
-\frac{\sqrt{6}C_{\mathcal{D}}}{4\pi r^3} \sum_{p=-2}^{2}
\left(-1\right)^{p}C^{\left(2\right)}_{-p}\left(\theta,\phi\right)\left[\hat{\boldsymbol{\mathcal{D}}}_1\otimes \hat{\boldsymbol{\mathcal{D}}}_2\right]^{\left(2\right)}_{p}\, ,
\eeq
where $\theta$ and $\phi$ are the polar and azimuthal angles between particle 1 and particle 2. 
We have defined the irreducible tensor product of two vector operators $\hat{\mathbf{A}}$ and $\hat{\mathbf{B}}$ as
\beq
\left[\hat{\mathbf{A}}\otimes \hat{\mathbf{B}}\right]^{\left(2\right)}_p&\equiv&
\sum_{m=-1}^{1} (-1)^p\sqrt{5}\tj{1}{m}{1}{p-m}{2}{-p}\hat{A}_m\hat{B}_{p-m}
\, ,
\eeq
Let us consider taking matrix elements of Eq.~(\ref{eq:HDDIv2}) for two magnetic dipoles.  
Using the basis $|FM_1M_2\rangle=|FM_1\rangle\otimes |FM_2\rangle$, we have
\beq
\label{eq:MDDIME}
\lefteqn{
\langle FM_1'M_2'|\hat{H}_{\mathrm{DDI};{\mu}}|FM_1M_2\rangle
}\\
&=&
-\frac{\sqrt{30}\mu_0}{4\pi r^3}\left(-1\right)^{2F-M_1'-M_2'}F\left(F+1\right)\left(2F+1\right)g_F^2\mu_B^2
\nonumber\\&&\times 
\sum_{p=-2}^2C^{\left(2\right)}_{-p}\left(\theta,\phi\right)
\sum_{m=-1}^{1} \tj{1}{m}{1}{p-m}{2}{-p} \tj{F}{-M_1'}{1}{m}{F}{M_1} \tj{F}{-M_2'}{1}{p-m}{F}{M_2}
\nonumber\, ,
\eeq
where we have used Eq.~(\ref{eq:totalMD}) to evaluate the dipole matrix elements.  
Similarly, for two symmetric top molecules in the same $(J,K)$-manifold 
we can evaluate the matrix elements in the basis $|JKM_1M_2\rangle=|JKM_1\rangle\otimes |JKM_2\rangle$, finding
\beq
\label{eq:EDDIME}
\lefteqn{
\langle JKM_1'M_2'|\hat{H}_{\mathrm{DDI};{d}}|JKM_1M_2\rangle
}\\
&=&
-\frac{\sqrt{30}}{4\pi\varepsilon_0 r^3}\left(-1\right)^{2J-M_1'-M_2'}d^2K^2\frac{2J+1}{J\left(J+1\right)}
\nonumber\\&&\times 
\sum_{p=-2}^2C^{\left(2\right)}_{-p}\left(\theta,\phi\right)
\sum_{m=-1}^{1} \tj{1}{m}{1}{p-m}{2}{-p} \tj{J}{-M_1'}{1}{m}{J}{M_1} \tj{J}{-M_2'}{1}{p-m}{J}{M_2}
\nonumber\, ,
\eeq
where we have used Eq.~(\ref{eq:totalED}) to evaluate the dipole matrix elements.  
Now, using the mapping Eq.~(\ref{eq:mapping1}), we find the mapping between the dipole-dipole matrix elements
\beq
\label{eq:mapping3}
\langle JKM_1'M_2'|\hat{H}_{\mathrm{DDI};{d}}|JKM_1M_2\rangle&\rightarrow&c^2\langle FM_1'M_2'|\hat{H}_{\mathrm{DDI};{\mu}}|FM_1M_2\rangle\, ,
\eeq
where $c^2=1/(\varepsilon_0\mu_0)$ is the speed of light in vacuum.  
The relations Eqs.~(\ref{eq:mapping1}), (\ref{eq:mapping2}), and (\ref{eq:mapping3}) establish a complete correspondence 
between the one- and two-body matrix elements of symmetric top molecules in a fixed $(J,K)$-manifold subject to a DC electric field 
and those of magnetic dipoles in a magnetic field, 
and hence between the many-body Hamiltonians of dilute gases of symmetric top molecules and magnetic dipoles.  
This correspondence is summarized in Table~\ref{table:mapping}, and may be seen visually in the special case of $J=1$ in Fig.~\ref{fig:singsigvsST}(c)-(f). 
{Appendix C} gives a complementary discussion in terms of the second quantized Hamiltonians.

\begin{table}
\caption{\label{table:mapping} 
Correspondence between symmetric top molecules in a fixed $(J,K)$-manifold subject to a DC electric field and magnetic dipoles in a magnetic field.  Note that the correspondence in Table~\ref{table:mapping} for the special case of a linear rigid rotor, i.e., $K=0$, gives a vanishing effective dipole moment.  That is to say, there is no such correspondence between magnetic dipoles and linear rigid rotors.
}
\begin{center}
\begin{tabular}{|c|c|c|}
\hline &Magnetic dipole & Symmetric top molecule\\
\hline Angular momentum &$F$&$J$\\
\hline Effective dipole moment &$g_F\mu_B$&$dK/J\left(J+1\right)$\\
\hline external field &$B$&$E_{\mathrm{DC}}$\\
\hline Dipole-dipole coupling coefficient & {$\mu_0$} &{$1/\varepsilon_0$} \\
\hline
\end{tabular}
\end{center}
\end{table}
 
To stress the usefulness of the mapping in Table~\ref{table:mapping}, we will discuss dimensional quantities.  
The mapping in Table~\ref{table:mapping} then tells us that 
the ratio of dipole-dipole interaction energies of a magnetic dipole of spin $F$ simulated with a symmetric top molecule 
and a true magnetic dipole of the same spin with $g$-factor $g_F$ is
\beq
\frac{E_{dd}^{d}}{E_{dd}^{\mu}}
&\simeq&
\frac{d_{\rm D}^2K^2}{g_F^2F^2(F+1)^2}\times 1.2\times 10^4
~,
\eeq
where $d_{\rm D}$ denotes the permanent electric dipole moment of the symmetric top molecule in units of Debye.  
For example, if we were to perform quantum simulation of $^{52}$Cr~\cite{Griesmaier_Werner_05} with {\MF} in the $J=K=3$ manifold, 
then the relevant parameters are $g_F=2$, $F=3$, 
and $d=1.85$, giving an enhancement of $\approx 620$ in the dipole-dipole interaction energy.  
Even for highly magnetic atoms, such as $^{168}$Er in the $F=6$ state~\cite{Er} and $^{164}$Dy in the $F=7$ state~\cite{DyBEC}, 
we see enhancements of $\approx 600$ and $\approx 310$, respectively.  
Thus, a quantum simulation of magnetic dipoles with symmetric top molecules allows us to access the physics of magnetic dipoles 
with much larger dipole-dipole interaction energies than are available with even the most magnetic atoms.

A few other notes regarding the mapping in Table~\ref{table:mapping} are in order.  
First, a very powerful feature of the mapping is that one can choose the effective spin of the simulated magnetic dipole 
by choosing the rotational state of the symmetric top molecule.  
Hence, while each atomic species has a fixed maximum angular momentum, one can choose in principle any effective spin degree of freedom 
using a symmetric top molecule.  
One restriction on the effective spin is that it be an integer, as the rotational quantum number $J$ is always an integer.   Furthermore, a practical restriction on the choice of $J$ is imposed by rotationally inelastic processes.   In this regard, states with $J=|K|$ are expected to be most stable, as there are no lower-lying states which have dipole-allowed transitions.  However, provided that rotational quenching rates are fast enough, states with $J>|K|$ may be stabilized in an optical lattice due to the quantum Zeno effect.  The quantum Zeno effect due to rapid chemical reaction rates has been demonstrated for KRb in an optical lattice~\cite{Yan_Moses}.  The restriction to integer spin $J$ does not imply that the simulated magnetic dipoles are bosons, as the effective spin of the simulated magnetic dipole is not associated with its statistics.  
Indeed, symmetric top molecules have hyperfine structure, as is discussed in {Sec.~\ref{sec:Hfs}}, 
and the hyperfine degrees of freedom will determine the statistics of the simulated magnetic dipole.  
Of the molecules in Table~\ref{table:SymmTop}, $^{13}${\MF} and $^{12}$CH$_{3}$CN are fermionic while the others are bosonic, 
but this list is by no means exhaustive.  
The fact that the effective spin and statistics of a symmetric top molecule do not have to be related opens the avenue of studying magnetic dipoles
 with integer angular momentum but fermionic statistics, something which does not occur in nature with elemental magnetic dipoles.   Some applications of our mapping for many-body physics will be discussed further in Sec.~\ref{sec:applications}.

\section{Applications of symmetric top molecules for many-body physics}
\label{sec:applications}

In this section, we discuss four applications of our correspondence between symmetric top molecules and magnetic dipoles for many-body physics, 
each of which can be called a quantum simulator. 
The first example is to simulate a gas of polarized dipoles. 
Symmetric top molecules can realize such a gas without demanding a large electric field as in ${}^1\Sigma$ molecules 
but with hundreds of times larger dipole-dipole interaction energies than those of magnetic atoms. 
Second, by loading a deep optical lattice with symmetric top molecules, 
we can obtain lattice spin models with long-range interactions. 
Any integer spin system is available by choosing the corresponding rotational $J$-manifold of symmetric top molecules. 
Third, simulation of molecular magnets is another example. Here, we suggest the possibility to simulate effective spin models for complex crystalline compounds with 
a gas of symmetric top molecules in an optical lattice where the effects of disorder are more readily controllable. 
Finally, we present an application for the multi-component physics in magnetic atoms.  In particular, we discuss the Einstein de-Haas effect and spin relaxation dynamics where the dipole-dipole interaction plays an important role and can be more clearly observed with the use of symmetric top molecules.

\subsection{Polarized dipoles}
\label{sec:polarized}
The simplest scenario of many-body physics with dipolar particles is a gas of single component, point particles 
which possesses resonant dipole moments.  Several novel physical phenomena occur for such a system, 
including the manifestation of a roton mode in a dipolar Bose-Einstein condensate~\cite{Goral_Shlyapnikov_03,Fischer}, 
BCS pairing in a dipolar degenerate Fermi gas~\cite{You_Marinescu,Baranov_Marenko}, 
crystalline phases in the two-dimensional strongly coupled limit~\cite{Kalia,Buechler_Demler,Astrakharchik}, 
supersolid~\cite{Goral_Santos,Kovrizhin,CGP,Pollet_Picon} and topologically ordered phases~\cite{DT_Berg} 
for dipolar gases in optical lattices, and vortices in rotating dipolar gases~\cite{Cooper_Rezayi,Komenias_Cooper,Zhang_Zhai,Yi_Pu,Baranov_Osterloh}, 
to name a few.  
More information about the physics of polarized dipolar gases may be found in a number of recent reviews of the subject~\cite{Baranov_08,Lahaye,Zoller_Rev}.

In order that a multicomponent particle realizes a single-component polarized dipole, 
three conditions must be met: 
(i) a single internal state is occupied, 
(ii) this state should possess a resonant dipole moment, 
and (iii) all other internal states should be separated by energies 
large compared to the characteristic dipole-dipole interaction energy such that state-changing collisions do not occur.  
For linear rigid rotors, the condition of a single internal state is easily satisfied by populating only a single level, say, 
the rotational ground state $|000\rangle$.  In the linear rigid rotor case, however, large electric fields must be applied to polarize a significant fraction of the dipole moment in the space-fixed frame, 
see Table~\ref{table:SingSig}.  When any rotor eigenstate has an appreciable resonant dipole moment due to application of a static electric field, 
all other dipole-allowed states lie naturally very far away in energy.  
For realizing the single-component polarized dipole with magnetic dipoles we assume that only the maximally stretched spin state, $|FM_F=F\rangle$, is populated, 
as this is the state with the largest dipole moment.  
Then, applying a magnetic field such that the Zeeman splitting between states $E_{\mathrm{field}}=g_F\mu_N B\gg E_{\mathrm{DD}}$, 
where $E_{\mathrm{DD}}$ is the characteristic dipole-dipole interaction energy, requirement (iii) is met.  
Often, the magnetic fields required are very small for magnetic dipoles.  
For example, in Chromium, a field on the order of 1 mG is sufficient~\cite{Pasquiou_Marechal_106}.  
For symmetric top molecules, we assume that only the $|JJJ\rangle$ level is populated, as it has the largest dipole moment.  
Applying an electric field such that $E_{\mathrm{field}}=dE_{\mathrm{DC}}\gg E_{\mathrm{DD}}$, we again satisfy requirement (iii).  
The fields required are on the order of V/m, much smaller than the kV/cm fields required to polarize linear rigid rotors.  
However, larger fields may be applied, and will result in larger effective dipole moments due to (usually small) Stark effects beyond first order.  
For $J=K=1$, already we obtain half of the resonant dipole moment.  
For higher $J=K$ states, we can obtain a larger fraction of the permanent dipole moment, see Eq.~(\ref{eq:dipmatelems}).  
Hence, we can obtain gases of single-component polarized dipoles with symmetric top molecules 
which interact orders of magnitude more strongly than magnetic dipoles and require orders of magnitude smaller electric fields than linear rigid rotors.

\subsection{{Lattice} spin models}
\label{sec:LSM}
Let us now consider the case in which a gas of symmetric top molecules 
in a specific $(J,K)$-manifold has been loaded into an optical lattice such that there exists exactly one molecule per lattice site.  We note that trapping of molecules in an optical lattice is demonstrated technology for $^1\Sigma$ alkali metal dimers; KRb has been trapped in an optical lattice~\cite{chotia_neyenhuis_11,Yan_Moses}, and RbCs is made directly in optical lattices~\cite{RbCs,lercherAD2011}, to name two examples.  In the limit that the lattice is very deep, which can be achieved by increasing the intensity of the optical lattice beams, all molecules reside 
in the lowest band of the lattice and the motional degrees of freedom are completely quenched.  
Hence, the dynamics of the system reduces to that of the dynamics of the internal degrees of freedom.  For simplicity, we assume that the tensor shift, Eq.~(\ref{eq:tensshift}), has been canceled, see Sec.~\ref{sec:otherfields}.  Then, the many-body dynamics of the internal degrees of freedom may be described, up to constant terms, by a lattice spin model 
\beq
\label{eq:Spinmodel}\hat{H}_{\mathrm{Spin}}&=&\frac{W}{2} \sum_{i\ne j}\sum_{p=-2}^{2}\sum_{m=-1}^1A^{p,m}_{i,j}\hat{S}^m_i\hat{S}^{p-m}_j-h\sum_i\hat{S}^0_i\, ,
\eeq
with
\beq
W&\equiv &\frac{d^2 }{4\pi \epsilon_0 a^3}\frac{K^2}{J^2\left(J+1\right)^2}\, ,\qquad h\equiv  \frac{d KE_{\mathrm{DC}}}{J\left(J+1\right)}\, ,\\
A^{p,m}_{i,j}&\equiv&-\sqrt{30}\tj{1}{m}{1}{p-m}{2}{-p}
\int {\rm d}\mathbf{r}{\rm d}\mathbf{r}' \left|w_i\left(\mathbf{r}\right)\right|^2
\frac{C^{\left(2\right)}_{-p}\left(\theta,\phi \right)}{\left|\mathbf{r}-\mathbf{r}'\right|^3}\left|w_j\left(\mathbf{r}'\right)\right|^2\, . 
\eeq
Here, $a$ is the lattice spacing, the functions $C^{\left(2\right)}_p\left(\theta,\phi\right)$ are the unnormalized spherical harmonics defined in Eq.~(\ref{eq:unnSH}), $w_i\left(\mathbf{r}\right)$ is a lowest band Wannier function~\cite{Kohn_59} centered on lattice site $i$, 
 and we have defined the {spin-$J$} operators of site $i$, $\hat{S}^p_i$, by the matrix elements
\beq
\langle JKM'|\hat{S}^p_i|JKM\rangle&=&\left(-1\right)^{J-M'}\tj{J}{-M'}{1}{p}{J}{M}\sqrt{J(J+1)(2J+1)}\, .
\eeq
The operators $\hat{S}^p_i$ mimic the behavior of a local spin of effective magnitude $S=J$ according to {Eqs.~(\ref{eq:magdips}) and (\ref{eq:magredmat}).}
The model Eq.~(\ref{eq:Spinmodel}) describes elemental spins of a length fixed by the rotation of the symmetric top molecule interacting 
{via} dipole-dipole interactions with characteristic coupling energy $W$ and placed in an effective magnetic field $h$.  
The coefficients $A^{p,m}_{i,j}$ describe the anisotropy of the process 
in which the net $Z$-component of the spin changes by $p$ as a function of the {distance} between sites $i$ and $j$.  
Neglecting effects due to anisotropic confinement~\cite{Wall_Carr_CE}, the spatial dependence of $A^{p,m}_{i,j}$ 
is given approximately as $C^{(2)}_{-p}(\theta_{ij},\phi_{ij})|i-j|^{-3}$, where $\mathbf{r}_i$ and $\mathbf{r}_j$ are 
the {spatial} coordinates of lattice sites $i$ and $j$, respectively, and $\theta_{ij}$ and $\phi_{ij}$ are the polar and azimuthal angles between lattice sites $i$ and $j$.  

A limiting case of the spin model Eq.~(\ref{eq:Spinmodel}) which is of particular interest is when the effective magnetic field $h$ is much larger than the spin-spin coupling $W$.  
In this case only the $p=0$ processes are {resonant}, 
and so the total magnetization $\langle \sum_i\hat{S}^0_i\rangle$ is conserved\footnote{Here, we neglect off-resonant processes in which field energy is exchanged for kinetic energy~\cite{Pasquiou_Bismut_106} and we also exclude the possibility that the field energy is near-resonant with an excitation energy to an excited band of the lattice~\cite{dePaz_Chotia_12}.}.  
Thus, the magnetic field term may be removed from the Hamiltonian by a gauge transformation and we are left with a long-range $XXZ$ model 
\beq
\label{eq:XXZ}\hat{H}_{\mathrm{XXZ}}&=&
\frac{W}{2}\sum_{i\ne j}\frac{1-3\cos^2\theta_{ij}}{\left|i-j\right|^3}\left\{\hat{S}^z_i\hat{S}^z_j-\frac{1}{4}\left[\hat{S}^+_i\hat{S}^-_j+\hat{S}^-_i\hat{S}^+_j\right]\right\}\, ,
\eeq
where we have switched from the spherical tensor notation for the spin operators to the more conventional ladder operator notation used in quantum magnetism
\beq
\hat{S}^z_i&=&\hat{S}^0_i\, ,\qquad
 \hat{S}^{\pm}_i~=~\hat{S}^x_i\pm i \hat{S}^y_i~=~\mp \sqrt{2}\hat{S}^{\pm 1}_i\, .
 \eeq
For {\MF} in a 532nm optical lattice, the coupling constant $W$ {becomes} on the order of {15 kHz} or larger, 
which is an order of magnitude larger than the corresponding coupling constant for the linear rigid rotor proposals with KRb 
and of the same order of magnitude as the coupling constant for LiCs in {the same proposals}~\cite{Gorshkov_Manmana_11}.  As this coupling constant can be on the order of the band gap, novel dipolar physics involving higher bands may also arise~\cite{Dutta_Sowinski}.  An important new feature of {our} XXZ model Eq.~(\ref{eq:XXZ}) compared to proposals using linear rigid rotors~\cite{Gorshkov_Manmana_11,Gorshkov_Manmana_11b, Hazzard_Manmana_13, Gorshkov_Hazzard_13,Micheli_Brennen_06,Brennen_Micheli_07,Barnett_Petrov_06,Manmana_Stoudenmire_13,Lemeshko_Krems} 
is that any {integral} spin $S$ can be achieved in principle without the fine-tuning of any fields by simply populating a different rotational manifold $J$.  
Systems with higher spin $S$ should more closely match the predictions of spin-wave theory, 
and so our results provide the possibility to assess the accuracy of approximate theoretical approaches~\cite{Peter_Mueller_12}.  
We expect that the addition of microwave fields to couple rotational states with the same $K$ can greatly extend the tunability of spin models such as Eq.~(\ref{eq:Spinmodel}), 
by analogy with what has been done for the case of linear rigid rotors~\cite{Gorshkov_Manmana_11,Gorshkov_Manmana_11b, Manmana_Stoudenmire_13, Gorshkov_Hazzard_13,Lemeshko_Krems}.  We leave a detailed discussion of microwave dressing of symmetric top molecules for future work.

\subsection{Molecular magnets}

One particular realization of large effective spins in condensed matter is crystals of molecular magnets.  Molecular magnets are organic molecules which contain transition metals with strongly exchange-coupled spins, and so each molecule behaves as a single large effective spin~\cite{Hill_MM}.  Prominent examples of molecular magnets are the manganese complex Mn$_{12}$O$_{12}$(CH$_3$COO)$_{16}$(H$_2$O)$_4$, referred to as {\Mn}~\cite{Mn,Mnac},  the manganese acetate complex Mn$_{12}$O$_{12}$(CH$_3$COO)$_{16}$(H$_2$O)$_4$$\cdot$CH$_3$COOH$\cdot$4H$_2$O, referred to as {\Mnac}, and the iron cluster [(tacn)$_6$Fe$_8$O$_2$(OH)$_12$]$^{8+}$, referred to as {\Fe}.  Here tacn is the cyclic organic ligand compound triazacyclononane with chemical formula C$_6$H$_{12}$(NH)$_3$.  All of the listed molecular magnets have effective spins $S=10$.  However, a range of other effective spins are possible, including half-integer spins~\cite{Hill_MM}.  Both {\Mn} and {\Fe} may be synthesized as large crystals; for example {\Mn} crystallizes in a tetragonal structure.  Additionally, as exemplified by {\Mnac}, {\Mn} and other molecular magnets can often be dissolved in a variety of solvents by ligand substitution, and this can lead to a variety of other crystalline structures.

Molecular magnetic crystals have been of great interest due to the possibility of observing resonant, coherent tunneling of magnetization.  In particular, the hysteresis loop describing the behavior of the magnetization as a magnetic field is decreased and then increased shows step-like features~\cite{Mn}.  These step-like features have been associated with transfer of magnetization between magnetic sublevels made degenerate by the applied magnetic field, either through thermally assisted~\cite{Mn} or purely quantum mechanical means~\cite{Wernsdorfer}.  In many treatments of crystals of molecular magnets, it is assumed that the lattice spacing is large enough to neglect dipole-dipole interactions between neighboring spins.  With this assumption, tunneling of magnetization occurs through terms such as the rhombic zero-field splitting term $E(\hat{S}_x^2-\hat{S}_y^2)$ which change the magnetization of a single molecule.  However, dipole-dipole interactions may indeed play a role in the tunneling of magnetization in molecular magnets, and can give rise to novel phenomena such as significant deviations from the predictions of Landau-Zener theory during a magnetic field quench~\cite{LZClusters}.  A quantum simulation of a crystal of molecular magnets with symmetric top molecules can feature much larger dipole-dipole interactions than in actual molecular magnets, and so deviations from Landau-Zener theory should be easier to see in such a system.

A quantum simulation of crystals of molecular magnets with symmetric top molecules is possible using a setup similar to Sec.~\ref{sec:LSM} with symmetric top molecules loaded into an optical lattice.  One advantage of using symmetric top molecules versus actual molecular magnets is the possibility of strong dipole-dipole coupling, as mentioned in the last paragraph.  We note that the dipole-dipole interaction can also be made weak while the effective spin is still large by choosing a small value for $K$ but a large value for $J$, see Eq.~(\ref{eq:dipmatelems}).  Another advantage, also discussed in Sec.~\ref{sec:LSM}, is the ability to change the effective spin $S$ of the simulated molecular magnet via the rotational state of the symmetric top molecules.  Furthermore, altering the optical lattice geometry and the filling allows for control over the crystalline geometry and disorder.  All of these additional controls may be useful in designing and optimizing molecular magnets for applications.  Finally, using the tensor shift of the optical lattice or microwave fields, it may be possible to simulate an axial zero-field splitting $D\hat{S}_z^2$ or the rhombic zero-field splitting mentioned above, both of which play a significant role in conventional molecular magnets.

\subsection{The Einstein-de Haas effect and spin relaxation}
\label{sec:edH}
Let us now consider a multi-component Bose-Einstein condensate (BEC) with dipole-dipole interactions \cite{Kawaguchi_2012}.  
The dipole-dipole interaction Eq.~(\ref{eq:HDDI1}) does not conserve the projection of the spin angular momentum, 
but conserves the projection of the total angular momentum, where the total angular momentum is the sum of internal spin and orbital angular momenta.
Hence, the process of the spin relaxation~\cite{Pasquiou_Marechal_106,Pasquiou_Marechal_108,dePaz_Chotia_12}, 
in which the projection of the total spin is altered via the dipole-dipole interaction, is associated with transfer of angular momentum
from the spin to the orbital angular momentum. 
This transfer of angular momentum means that an initially polarized gas of dipoles starts to rotate spontaneously, 
a phenomenon known as the Einstein-de Haas effect~\cite{EdH_1915}
, possibly producing non-singular vortices in the multi component BEC. 
Based on mean field theories~\cite{Kawaguchi_2006,Santos_2006}, 
the dynamics of spin relaxation is interpreted as 
the Larmor precession of dipoles around the effective magnetic field produced 
by inhomogeneous condensates of bosonic dipoles.  
Thus, the dipole-dipole interaction determines a typical time scale for the spin relaxation, 
which for bosonic atoms is on the order of $\tau_{\mathrm{sr}}\simeq h/(c_{dd}n)$ 
where $n$ is the density and $c_{dd}=\mu_0(g_F\mu_B)^2/(4\pi)$. 
For a ${}^{52}$Cr BEC with $n\sim 10^{14}{\rm cm}^{-3}$, 
$\tau_{\mathrm{sr}}$ is a fraction of a millisecond, 
while for symmetric top molecules with $J=K=3$ it can be shortened to the order of microseconds for the same density
due to a six hundred-fold increase of $c_{dd}$ with respect to ${}^{52}$Cr, see Sec.~\ref{sec:mapping}.  
Likewise, spin relaxation dynamics for symmetric top molecules can be seen on the same timescales as for Chromium 
with densities a factor of six hundred smaller.  In addition, in order to see spin relaxation, 
the characteristic energy difference between single-particle states with neighboring spin projections 
$M$ and $M\pm 1$ should be small compared to the characteristic dipole-dipole interaction energy.  
For ${}^{52}$Cr, the dipole-dipole interaction is weak, and so spin relaxation requires very small magnetic fields of 1~mG or less~\cite{Ho_2006}.  
For symmetric top molecules, as discussed in Sec.~\ref{sec:Stark}, 
the energetic differences between states with neighboring projections $M$ and $M\pm 1$ can be small compared 
with the dipole-dipole interaction for a wide range of the applied DC electric field. There are many other studies on the physics of multi-component magnetically dipolar atoms, 
such as spin textures~\cite{Yi_2006,Kawaguchi_2007,Vengalattore_2008} and dipolar relaxation~\cite{Pasquiou_Bismut_81,Pasquiou_Bismut_106}; 
we expect that quantum simulation with symmetric top molecules can access these phenomena on shorter timescales or with lower density 
and even explore new regimes with large dipole-dipole interaction energies compared to isotropic contact interaction energies.

\section{The influence of hyperfine structure}
\label{sec:Hfs}

The correspondence between symmetric top molecules and magnetic dipoles has been discussed within the context of the rigid rotor approximation (RRA), 
where the nuclei are structureless objects rigidly connected in space.  
Within this approximation, the states $|JKM\rangle$ and $|J,-KM\rangle$ are degenerate due to the cylindrical symmetry of the molecule, 
and possess resonant {space-fixed dipole moments} $\langle JKM|\hat{d}_0|JKM\rangle=dKM/[J(J+1)]$ even in the absence of an electric field.  
As mentioned in the introduction, the existence of an elemental electric dipole moment 
would violate both parity and time reversal symmetries~\cite{withoutStrangeness}.  
While neither parity nor time reversal are good symmetries of the universe within the context of the standard model, 
their violations are extraordinarily slight~\cite{electronEDM,NeutronEDM}.  
Thus, it is expected that the existence of {space-fixed dipole moments} for symmetric top molecules in the RRA is a defect of the chosen approximate basis 
rather than being due to breaking of near-fundamental symmetries of the universe.  Any interaction which couples the states $|JKM\rangle$ and $|J,-KM\rangle$ will result in the proper physical eigenstates in the absence of fields being instead the linear combinations $|JKM\rangle\pm |J,-KM\rangle$, and introduce some energetic splitting $\Delta_{JKM}$ between these states~\cite{Klemperer_Lehmann,Lemeshko_Friedrich}.  These linear combinations have no expected dipole moment, and hence will not display a linear Stark effect.  The transition to the basis in which $|JKM\rangle$ and $|J,-KM\rangle$ are decoupled and hence a linear Stark effect is observed occurs for electric field strengths of order $\Delta_{JKM}/d$.  In this section, we will show that the splittings $\Delta_{JKM}$ are on the order of a few kHz for the $|K|=1$ levels and un-observably small for all higher $|K|$ levels.  Hence, the linear Stark effect holds for essentially any non-vanishing field for $|K|>1$, and for fields greater than V/m for $|K|=1$.  The fields where the linear Stark effect begins to break down are many orders of magnitude larger than this, see Table~\ref{table:SymmTop}.  In this sense, accounting for terms which couple together states with $\pm |K|$ does not significantly affect the mapping of Sec.~\ref{sec:Reln}. 

For molecules with $C_{3v}$ symmetry\footnote{A brief review of molecular point group symmetry is given in Sec.~\ref{sec:MolSymm}.} and non-zero nuclear spin such as {\MF}, states with $|K|\ne 3N$ ($N$ an integer) are degenerate to all orders in the rotation-vibration Hamiltonian.  Thus, breaking of the $\pm|K|$ degeneracy dominantly occurs through hyperfine effects for such molecules.  In the following subsections we will investigate the role of hyperfine structure in the rotational structure of the canonical symmetric top molecule {\MF}, shown in Fig.~\ref{fig:Schematic}.  A more general discussion of the breaking of the $\pm|K|$ degeneracy for non-planar symmetric tops of the form XY$_3$, including symmetric tops with no nuclear spin, may be found in Ref.~\cite{Klemperer_Lehmann}.  The dominant hyperfine interactions of symmetric top molecules without nuclear quadrupole coupling are spin-rotation interactions and spin-spin interactions.  We discuss these interactions in Secs.~\ref{sec:SR} and \ref{sec:SS}, respectively.  Before discussing these interactions, we briefly review the classification of molecular levels according to point group and exchange symmetries of constituent nuclei in Sec.~\ref{sec:MolSymm} for non-specialists.  Finally, in Sec.~\ref{sec:NZ}, we discuss the nuclear Zeeman effect.  In addition to discussing hyperfine interactions for the purpose of clarifying the degree of breaking of the $\pm|K|$ degeneracy, isolation of particular hyperfine states is often an important step in reaching quantum degeneracy~\cite{Hfs}, and so we compile information about the hyperfine structure of {\MF} here to aid future research on the many-body physics of symmetric top molecules.

\subsection{Classification of molecular symmetry}
\label{sec:MolSymm}
A classification of the symmetry of a molecule using the language of point groups 
results in the full permutation-inversion group of the molecule~\cite{Jensen_Bunker} 
in which the eigenstates of the molecule transform irreducibly under spatial inversions as well as exchanges of identical nuclei.  
Using group theory, it can be shown that the only states allowed by exchange symmetry belong to one of two non-degenerate representations 
which have opposite parity~\cite{Watson_FOSE,Jensen_Bunker}.  
For {\MF}, the molecular symmetry group is $C_{3v}$, which corresponds to a 3-fold rotation axis which exchanges the three Hydrogen nuclei in a clockwise fashion 
as well as reflections through any plane which contains the symmetry axis and one of the Hydrogen nuclei.  
There exist three representations of $C_{3v}$, denoted $A_1$, $A_2$, and $E$.  
The representation $A_1$ is completely symmetric under all group operations, and both $A_1$ and $A_2$ are non-degenerate.  
The representation $E$ is doubly degenerate.  In addition to the rotational wavefunctions, 
the three Hydrogen nuclear spins can couple to form either ortho-{\MF} ($I^{\mathrm{H}}=3/2$) or para-{\MF} ($I^{\mathrm{H}}=1/2$), with $I^{\mathrm{H}}$ the total nuclear spin of the H nuclei.  
A detailed examination of the molecular symmetry group, taking into account indistinguishability of the Hydrogen nuclei, 
demonstrates that not all combinations of nuclear and rovibrational wave functions give internal wave functions with the appropriate symmetry.  
In particular, when $K=3N$, $N$ a positive integer, the rovibrational wave function has symmetry $A_1\oplus A_2$, 
and so the nuclear spin function must belong to the completely symmetric representation $A_1$, ortho-{\MF}, 
in order that the internal wave function has the correct symmetry.  
In contrast, when $K=3N\pm 1$ the rovibrational wave function has $E$ symmetry, 
and so the nuclear spin wave function must also have $E$ symmetry, para-{\MF}, in order that the complete wave function transforms properly.  
The actual rovibrational and spin wavefunctions corresponding to these representations have been given 
in the literature~\cite{Townes_Schawlow,Jensen_Bunker,Chapovsky,Harder_Macholl}, 
and the methods of group theory can be applied to simplify evaluation of matrix elements, just as for the case of SU(2)~\cite{Recoupling}.

\subsection{Spin-rotation interactions}
\label{sec:SR}
Nuclear spin-rotation interactions may be written as
\beq
\hat{H}_{\mathrm{s-r}}&=&-\sum_X \sum_{i} \, \hat{\mathbf{I}}^{X_{(i)}}\cdot \tilde{c}^{X_{(i)}}\cdot \hat{\mathbf{J}}\, ,
\eeq
where $X$ is a label for nuclei, e.g., $X=\{$H,C,F$\}$ for {\MF}, 
and $i$ runs over the different members of that nucleus, e.g., H$_{(i)}$ takes H$_{(1)},$ H$_{(2)}, $ H$_{(3)}$ for {\MF}.  
The tensor element $\tilde{c}^{X_{(i)}}_{\alpha\beta}$ expresses the magnetic field in the $\alpha^{\mathrm{th}}$ direction at nucleus $X_{(i)}$ 
due to rotation of the molecule about direction $\beta$.  
For a nucleus $X$ which lies on the symmetry axis, the principal axes of the tensor $\tilde{c}^X$ are parallel to those of the inertia tensor, 
and $\tilde{c}^X$ also has a doubly degenerate eigenvalue.  
The matrix elements can be computed by analogy with the rotational Zeeman effect, Eq.~(\ref{eq:HrotZ}), 
with  the operators $\hat{{I}}^{X_{(i)}}_p$ taking the place of magnetic field projections $B_p$.  
Hence, as with the rotational Zeeman effect, the spin-rotation interaction for nuclei on the symmetry axis will not mix states with different values of $K$.  
For {\MF}, the values {$c^{\mathrm{F}}_{\perp}=4.0$kHz} and {$c^{\mathrm{F}}_{\parallel}=-51.1$kHz} have been determined experimentally~\cite{Wofsy}, 
and {$c^{\mathrm{C}}_{\perp}=-6.03$kHz and $c^{\mathrm{C}}_{\parallel}=-18.33$kHz} have been computed for $^{13}$C ($^{12}$C has zero nuclear spin)~\cite{conv}.  

Let us now consider nuclei which do not lie on the symmetry axis, such as the H atoms in {\MF}.  
If we define the coordinate system such that the molecule-fixed $x$-axis points from the symmetry axis to the Hydrogen nucleus labeled 1, 
then reflection symmetry about the plane passing through the symmetry axis and this Hydrogen nucleus dictates that the spin-rotation tensor $\tilde{c}^{\mathrm{H}_{(1)}}$ 
takes the general form
\beq
\tilde{c}^{\mathrm{H}_{(1)}}&=&\left(\begin{array}{ccc} c_{xx}^{\mathrm{H}}&0&c_{xz}^{\mathrm{H}}\\ 0&c_{yy}^{\mathrm{H}}&0\\ c_{zx}^{\mathrm{H}}&0&c_{zz}^{\mathrm{H}}\end{array}\right)\, ,
\eeq
in the molecule-fixed frame.  The tensor $\tilde{c}^{{\mathrm{H}}_{(1)}}$ is not \emph{a priori} symmetric when computed in a specific reference frame, 
and so the spin-rotation Hamiltonian of the first Hydrogen nucleus must be made manifestly Hermitian{~\cite{Harder_Macholl} as} 
$\hat{H}_{\mathrm{s-r};{\mathrm{H}}_{(1)}}\to
 \frac{1}{2}\left[\hat{H}_{\mathrm{s-r};{\mathrm{H}}_{(1)}}+\hat{H}_{\mathrm{s-r};{\mathrm{H}}_{(1)}}^{\dagger}\right]$.  
 This amounts to replacing $c_{xz}^{\mathrm{H}}$ with $\left(c_{xz}^{\mathrm{H}}+{c_{zx}^{\mathrm{H}}}^{\star}\right)/2$ and $c_{zx}^{\mathrm{H}}=\left(c_{zx}^{\mathrm{H}}+{c_{xz}^{\mathrm{H}}}^{\star}\right)/2$.  
Because of the three-fold rotational symmetry of the molecule, 
we have that $\tilde{c}^{{\mathrm{H}}_{(2)}}=\tilde{R}_2^{-1}\tilde{c}^{{\mathrm{H}}_{(1)}}\tilde{R}_2\,$ and $\tilde{c}^{{\mathrm{H}}_{(3)}}=\tilde{R}_3^{-1}\tilde{c}^{{\mathrm{H}}_{(1)}}\tilde{R}_3$, 
where
\beq
\tilde{R}_2&=&\tilde{R}_3^{-1}~=~
\begin{pmatrix}
-1/2 & \sqrt{3}/2 & 0 \\
-\sqrt{3}/2 & -1/2 & 0 \\
0 & 0 & 1
\end{pmatrix}
\eeq
is the matrix which rotates the molecule by $2\pi/3$ about the symmetry axis.  
The values of the Hydrogen nuclear spin-rotation tensor have been calculated to be 
$c_{xx}^{{\mathrm{H}}}=-1.25$, $c_{yy}^{{\mathrm{H}}}=2.59$, $c_{zz}^{{\mathrm{H}}}=15.92$, $c_{xz}^{{\mathrm{H}}}=5.77$, $c_{zx}^{{\mathrm{H}}}=1.41$ for $^{12}${\MF}~\cite{conv}, 
with deviations of a few percent for $^{13}${\MF}~\cite{conv}.  
As opposed to the case of spin-rotation interactions for nuclei on the symmetry axis, 
the transformation of the off-axis spin-rotation interaction to the space-fixed frame will not result in a summation of spherical harmonics.  
Rather, the summation will involve rotation matrices $\mathscr{D}^{j}_{\Delta M\Delta K}(\omega)$ with both $\Delta M$ and $\Delta K$ nonzero.  
In particular, this implies that such an interaction can mix levels whose values of $K$ differ by at most two, as the interaction involves at most rank-two tensors.  
A physical way to interpret this result is that the Hydrogen nuclear spin-rotation interaction breaks the cylindrical symmetry of the molecular Hamiltonian, 
as the Hydrogen nuclear spins have only discrete rotational symmetry about the symmetry axis.  
The only pair of projections $K$ which are degenerate and connected by $\Delta K=\pm 2$ are the $|K|=1$ states.  
Here, the Hydrogen spin-rotation interaction splits the symmetric and antisymmetric linear combinations of $|J1M\rangle$ and $|J,-1M\rangle$ by a few kHz, 
and the linear behavior of the Stark effect does not set in until the field strength becomes comparable to this splitting.  
For states with $|K|>1$, the spin-rotation coupling only acts in second and higher order.  
The resulting splittings between the symmetric and antisymmetric linear combinations of $|JKM\rangle$ and $|J,-KM\rangle$ 
are of the order of $10^{-4}$Hz or less, and further decrease exponentially with increasing $|K|$~\cite{Klemperer_Lehmann}.

\subsection{Nuclear spin-nuclear spin interactions}
\label{sec:SS}
Neglecting any electron-mediated interactions between nuclear spins~\cite{Ramsey,Brown_Carrington,Ozier_Meerts_81,WofsyCH3D}, 
the nuclear spin-nuclear spin Hamiltonian contains only dipole-dipole interactions between all non-zero nuclear spins in the molecule.  It may be written as 
\beq
\label{eq:Hss}
\hat{H}_{\mathrm{s-s}}&=&-\frac{\sqrt{6}\mu_0\mu_N^2}{4\pi }
\sum_{<X_{(i)},X'_{(i')}>}\frac{g_Xg_{X'}}{R_{X_{(i)}X'_{(i')}}^3} 
\left[\hat{\mathbf{I}}^{X_{(i)}}\otimes \hat{\mathbf{I}}^{X'_{(i')}}\right]^{\left(2\right)}\cdot 
\left[\mathbf{e}_{X_{(i)}X'_{(i')}}\otimes \mathbf{e}_{X_{(i)}X'_{(i')}}\right]^{\left(2\right)}\, ,\qquad
\eeq
where 
the notation $<X_{(i)},X'_{(i')}>$ denotes a non-ordered pair of $X_{(i)}$ and $X'_{(i')}$, that is, each pair of nuclei is counted exactly once, 
$R_{X_{(i)}X'_{(i')}}$ is the distance between the nuclei $X_{(i)}$ and $X'_{(i')}$, 
and $\mathbf{e}_{X_{(i)}X'_{(i')}}$ is a unit vector pointing from nucleus $X_{(i)}$ to nucleus $X'_{(i')}$.  
The nuclear $g$-factors $g_{X}$ are well-known and tabulated~\cite{GreenBook}.  
For convenience, we note here that the $g$-factors relevant for {\MF} are $g_{\mathrm{H}}=5.585694$, $g_{\mathrm{F}}=5.257736$, and $g_{\mathrm{C}}=1.404822$ for $^{13}$C.  
Spin-spin interactions between the Hydrogen nuclei only contribute when the ortho configuration of nuclear spin is involved, 
and hence are only present when $K=3N$ {with a positive integer $N$}.  
On the other hand, the Fluorine-Hydrogen spin-spin interactions contribute for any configuration of the Hydrogen nuclear spin.  
Similarly to the Hydrogen nuclear spin-rotation interaction, 
the Fluorine-Hydrogen spin-spin interaction involves rotation matrices $\mathscr{D}^{j}_{\Delta M\Delta K}(\omega)$ 
when expressed in the space-fixed frame, and so it mixes states with $K$ values differing by at most $\Delta K=\pm 2$.  
As discussed in the above, this interaction causes splittings of a few kHz for the $|K|=1$ levels~\cite{WofsyCH3D}, 
but does not have observable consequences for higher values of $K$.  
Hence, we conclude that the $\pm K$ degeneracy is broken slightly for the $|K|=1$ levels, 
but can be considered to hold for all other levels $|K|>1$.  
In turn, this means that the linear Stark effect which forms the basis of our correspondence between symmetric top molecules and magnetic dipoles 
also holds for these states even in the presence of hyperfine structure.

\subsection{The nuclear Zeeman effect}
\label{sec:NZ}
The hyperfine interaction we lastly consider is the nuclear Zeeman effect, with Hamiltonian
\beq
\hat{H}_{\mathrm{Z;nuc}}&=&-\mu_N \sum_{X}\sum_{i}g_{X}\hat{\mathbf{I}}^{X_{(i)}}  \cdot \mathbf{B}\, .
\eeq
The $g$-factors were listed following Eq.~(\ref{eq:Hss}) above.  
For a magnetic field aligned along $\hat{\bf{e}}_Z$, the matrix elements are completely diagonal in the space-fixed frame:
\beq
\langle JKMI^{\mathrm{H}}M_{\mathrm{H}}M_{\mathrm{C}}M_{\mathrm{F}}|\hat{H}_{\mathrm{Z;nuc}}|JKMI^{\mathrm{H}}M_{\mathrm{H}}M_{\mathrm{C}}M_{\mathrm{F}}\rangle&=&-\mu_N\left(g_{\mathrm{H}}M_{\mathrm{H}}+g_{\mathrm{C}}M_{\mathrm{C}}+g_{\mathrm{F}}M_{\mathrm{F}}\right)B\, ,
\qquad
\eeq
where $M_{\mathrm{H}}$ is the sum of nuclear spin projections of Hydrogens.  
This expression holds for all representations of the Hydrogen nuclear wavefunction.

\section{Conclusions}
\label{sec:concl}

We have shown that the many-body physics of symmetric top molecules in states with nonzero projection of the molecular rotation along the symmetry axis maps onto the many-body physics of magnetic dipoles with effective spin set by the rotational quantum number of the molecule.  The strength of dipole-dipole interactions in the simulated magnetic dipoles is typically two orders of magnitude larger than for naturally occurring magnetic dipoles, which opens up new regimes of dipole-dipole interacting systems.  We illustrated several manifestations of novel many-body physics and quantum simulation hence possible with symmetric top molecules.  
In one example, symmetric top molecules can generate a gas of single-component polarized dipoles 
with either bosonic or fermionic statistics.  
The preparation of such a gas requires much smaller static electric fields than a polarized gas of linear rigid rotors, 
and will feature larger dipole-dipole interactions than a polarized gas of magnetic dipoles.  
Furthermore, we showed that symmetric top molecules in a deep optical lattice may be used 
to engineer quantum simulations of lattice spin models with anisotropic, long-range, and magnetization-changing interactions 
as well as arbitrarily large spin.  
Trapping of symmetric top molecules in an optical lattice also allows for quantum simulation of molecular magnets, 
where symmetric top molecules offer the advantage over condensed matter realizations of tuning the geometry, 
disorder, strength of interactions, and effective spin.  
Finally, we discussed quantum simulation of spontaneous demagnetization 
and the Einstein-de Haas effect on much shorter timescales or lower densities than with naturally occurring magnetic dipoles.  The existence of $K\ne 0$ states is a feature of symmetric top molecules which is not shared by linear rigid rotors, 
such as the $^1\Sigma$ ground states of the alkali dimers.  In addition to the correspondence between symmetric top molecules and magnetic dipoles, we also discussed some technologically important aspects of symmetric top molecules, such as opto-electrical cooling to reach quantum degeneracy, optical trapping, 
and the effects of hyperfine structure.

Our work has just begun the study of many-body physics with symmetric top molecules.  
In particular, we have only studied the cases of static or far-off-resonant fields.  
We expect that the use of microwave dressing will greatly enhance the tunability of effective many-body models 
with symmetric top molecules, just as has been the case for linear rigid rotors~\cite{Gorshkov_Manmana_11,Gorshkov_Manmana_11b}.  
In addition, threshold scattering of symmetric top molecules at ultralow temperatures is an unexplored area 
{and} beyond the scope of the present paper.  
Studies of ultracold scattering will provide insight into the relative role of short-range contact interactions 
to the long-range dipole-dipole interactions considered in this paper.

\begin{acknowledgement}
We thank Yehuda Band, Bretislav Friedrich, Alexey Gorshkov, Jeremy Hutson, Bruno Laburthe-Tolra, Biao Wu, and Martin Zeppenfeld for useful discussions.  This research was  supported in part by the National Science Foundation under Grants PHY-1207881 and NSF PHY11-25915 and by AFOSR grant number FA9550-11-1-0224.  We would like to acknowledge KITP for hospitality.
\end{acknowledgement}

\section{Appendix A: Matrix elements for symmetric top molecules}
\label{sec:AppendixA}

In this section we compile useful matrix elements for symmetric top molecules.  
As the wave functions of symmetric top molecules are proportional to Wigner $D$-matrices, see Eq.~(\ref{eq:STwfcns}), 
most matrix elements can be computed using 
the theorem on integration over products of three rotation matrices
~\cite{Zare_1988}
\beq
\int d\omega \mathscr{D}^{j_1}_{m_1'm_1}\left(\omega\right)\mathscr{D}^{j_2}_{m_2'm_2}\left(\omega\right)\mathscr{D}^{j_3}_{m_3'm_3}\left(\omega\right)
&=&8\pi^2\tj{j_1}{m_1'}{j_2}{m_2'}{j_3}{m_3'}\tj{j_1}{m_1}{j_2}{m_2}{j_3}{m_3}\, .\qquad
\eeq
In particular, using the facts that $\mathscr{D}^{j}_{m 0}\left(\omega\right)=C^{\left(j\right)\ast}_m\left(\omega\right)$ 
and $\mathscr{D}^{j{\ast}}_{m m'}=\left(-1\right)^{m-m'}\mathscr{D}^{j}_{-m -m'}$, we find for the expectation of any unnormalized spherical harmonic
\beq
\lefteqn{
\langle J'K'M'|C^{\left(\ell\right)}_p\left(\omega\right)|JKM\rangle
}\nonumber\\
&=&
\left(-1\right)^{M'-K}\sqrt{\left(2J'+1\right)\left(2J+1\right)}\tj{J'}{-M'}{\ell}{p}{J}{M}\tj{J'}{-K}{\ell}{0}{J}{K}\delta_{K'K}\, .
\qquad
\label{eq:MEC}
\eeq
As for the expectation of the dipole operator along a space-fixed direction $\mathbf{e}_p$, 
we note that the permanent dipole moment of a molecule must lie along the symmetry axis.  
We define the symmetry axis to be the molecule-fixed {$z$-axis} $\mathbf{e}_z=\mathbf{e}_{q=0}$, 
which yields ${\hat{\mathbf{d}}}=d\,\mathbf{e}_{q=0}$.  
We now rotate from the molecule-fixed frame to the space-fixed frame as~\cite{Brown_Carrington}
\beq
\hat{d}_p&=&\sum_{q=-1}^{1} \mathscr{D}^{1\ast}_{pq}\, \hat{d}_q=d\,C^{\left(1\right)}_0\, .
\eeq
Hence, the expected dipole moment along space-fixed direction $\mathbf{e}_p$ is given by Eq.~(\ref{eq:MEC}) as
\beq
\lefteqn{
\langle J'K'M'|\hat{d}_p|JKM\rangle
}\nonumber\\
&=&\left(-1\right)^{M'-K}\sqrt{\left(2J'+1\right)\left(2J+1\right)}\tj{J'}{-M'}{1}{p}{J}{M}\tj{J'}{-K}{1}{0}{J}{K}\delta_{K'K}\, .
\qquad
\eeq

\section{Appendix B: Microscopic angular momenta and effective angular momentum of a magnetic dipole}
\label{sec:angmomrel}

In this section, we discuss the 
relationship between the algebraic structure of the intrinsic angular momenta of an atom 
and the effective magnetic moment associated with the total angular momentum of the atom. 
Let us consider an atom with orbital angular momentum $L$, spin angular momentum $S$ of electrons, 
and nuclear spin angular momentum $I$ of nuclei.  
The magnetic dipole operator of the atom is
\beq
\label{eq:magDoperator}\hat{\boldsymbol{\mu}}&=&\mu_B\left(g_l\hat{\mathbf{L}}+g_e\hat{\mathbf{S}}\right)+\mu_Ng_N\hat{\mathbf{I}}\, .
\eeq
Now, we consider the fully coupled representation 
in which we first couple $L$ and $S$ to form $J$, $\hat{\mathbf{J}}=\hat{\mathbf{L}}+\hat{\mathbf{S}}$, 
and then couple $\hat{\mathbf{J}}$ and $\hat{\mathbf{I}}$ to form the total angular momentum $\hat{\mathbf{F}}$, $\hat{\mathbf{F}}=\hat{\mathbf{J}}+\hat{\mathbf{I}}$.  
The associated basis is denoted by $\{\ket{((L,S)J,I)FM_F}\}$.  
Using standard angular momentum decoupling techniques \cite{Zare_1988}, 
we can compute the projections of the intrinsic angular momenta in this basis as
\beq
\lefteqn{
\bra{((L',S')J',I')F'M'_{F'}}
\hat{L}_p
\ket{((L,S)J,I)FM_F}
}\nonumber\\
&=&
\delta_{LL'}\delta_{SS'}\delta_{II'}
\bra{F'M'_{F'}}\hat{F}_p\ket{FM_F}
\nonumber\\
&& \times
\frac{\left[
F(F+1)+J(J+1)-I(I+1)
\right]}
{2F(F+1)}
\frac{\left[
J(J+1)+L(L+1)-S(S+1)
\right]}
{2J(J+1)}
~,
\\
\lefteqn{
\bra{((L',S')J',I')F'M'_{F'}}
\hat{S}_p
\ket{((L,S)J,I)FM_F}
}\nonumber\\
&=&
\delta_{LL'}\delta_{SS'}\delta_{II'}
\bra{F'M'_{F'}}\hat{F}_p\ket{FM_F}
\nonumber\\
&& \times
\frac{\left[
F(F+1)+J(J+1)-I(I+1)
\right]}
{2F(F+1)}
\frac{\left[
J(J+1)+S(S+1)-L(L+1)
\right]}
{2J(J+1)}
~,
\\
\lefteqn{
\bra{((L',S')J',I')F'M'_{F'}}
\hat{I}_p
\ket{((L,S)J,I)FM_F}
}\nonumber\\
&=&
\delta_{LL'}\delta_{SS'}\delta_{II'}
\bra{F'M'_{F'}}\hat{F}_p\ket{FM_F}
\frac{\left[F(F+1)+I(I+1)-J(J+1)\right]}{2F(F+1)}
~. 
\eeq
Importantly, we see that all three projections are proportional to the projection of the total angular momentum $\hat{\mathbf{F}}$ 
as a result of the fact that all intrinsic angular momenta are spherical tensors.  
Hence, we obtain for the projection of the magnetic dipole operator, Eq.~(\ref{eq:magDoperator}),
\beq
\label{eq:effMag}
\bra{((L',S')J',I')F'M'_{F'}}
\hat{\mu}_p
\ket{((L,S)J,I)FM_F}
&=&
  g_F \mu_B
\bra{F'M'_{F'}}\hat{F}_p\ket{FM_F}
\, \delta_{LL'}\delta_{SS'}\delta_{II'}
~,
\qquad
\eeq
where the effective g-factor, or generalized Land\'{e} g-factor, is given by
\beq
g_{F}
&=&
C(F,J,I)
\left[ g_l C(J,L,S)+g_e C(J,S,L)\right]
+ \frac{\mu_N}{\mu_B} g_NC(F,I,J)
~,
\label{eq:gfactor}
\eeq
with a function 
\beq
C(\alpha, \beta, \gamma)
&=&
\frac{\alpha(\alpha+1)+\beta(\beta+1)-\gamma(\gamma+1)}{2\alpha(\alpha+1)}
~.
\eeq
Since 
$\mu_N/\mu_B=m_e/m_p \simeq 5\times 10^{-4}$, the last term in the braces may be neglected when considering the overall magnitude of the magnetic dipole.  
However, it is important to note that the nuclear spin implicitly contributes to the magnetic dipole moment via the total angular momentum.  
Also, substituting $g_l=1$ and $g_e\simeq 2$, we have
\beq
\bra{((L',S')J',I')F'M'_{F'}}
\hat{\mu}_p
\ket{((L,S)J,I)FM_F}
&\simeq&
\tilde{g}_F \mu_B 
\bra{F'M'_{F'}}\hat{F}_p\ket{FM_F}
\,\delta_{LL'}\delta_{SS'}\delta_{II'}
~, 
\label{eq:effMagapp}
\qquad
\eeq
with the approximate value for $g_F$, 
\beq
\tilde{g}_F
&=&
C(F,J,I)
\left[ 1+ C(J,S,L)\right]
~.
\eeq
When we are focusing on a subspace with fixed values of $L$, $S$, $J$, $I$, and $F$, 
which is denoted simply by $\mathscr{L}(\{\ket{FM_F}\}_{M_F=-F}^F)$, 
the effective g-factor Eq.~(\ref{eq:gfactor}) becomes constant, 
and thus the magnetic dipole moment $\hat{\boldsymbol{\mu}}$ is proportional to to the total angular momentum $\hat{\bf F}$ 
as a linear operator in this subspace, 
\beq
\hat{\boldsymbol{\mu}}&=&g_F\mu_B\hat{\mathbf{F}}
\qquad {\rm in}\quad  \mathscr{L}(\{\ket{FM_F}\}_{M_F=-F}^F)
\label{eq:mutoF}
\, .
\eeq 
Note that such a subspace spans an effective Hilbert space in several important cases 
and that it is useful to employ the equality~(\ref{eq:mutoF}).  
The first case is rather trivial where only a single angular momentum $\hat{\bf L}$, $\hat{\bf S}$, or $\hat{\bf I}$ is present, 
as then Eq.~(\ref{eq:mutoF}) becomes true in a whole Hilbert space.  
This is the case for Chromium, which contains only electronic spin in its bosonic isotopes~\cite{Griesmaier_Werner_05}.  
In addition, the effective theory holds whenever $I$ is zero and there is large separation between $F$-manifolds 
compared to the Zeeman splittings within an $F$-manifold.  Such a splitting comes from, e.~g., spin-orbit coupling.  
All of the bosonic isotopes of the most highly magnetic atoms Dy and Er have zero nuclear spin due to the fact that they 
have an even number of both protons and neutrons~\cite{NucPhys}, a trend which holds for a large majority of the highly magnetic atoms.  
The splittings between $F$-manifolds are many orders of magnitude larger than 
any possible inter-manifold Zeeman splittings in these atoms~\cite{SeoHeoThesis,DyBEC,Er}, and so the effective therory works.  
Finally, the effective description holds whenever all three angular momenta are present, but the spacings between hyperfine manifolds are large.  
This is the case for the fermionic isotopes of Dy and Er, as they all have large nuclear spins and, accordingly, 
large nuclear quadrupole interactions on the order of hundreds of MHz~\cite{DyHyperfine,DyFermi,SeoHeoThesis,ErHfs}.  
The hyperfine spin-spin interactions in the bosonic isotope $^{169}$Tm~\cite{Tm} are similarly large, despite the fact that there is no nuclear quadrupole~\cite{TmHyperfine}.  
The description of the magnetic dipole in terms of the total angular momentum 
hence holds up to magnetic fields which are much larger than the fields required to reach the regime of polarized dipoles, see Sec.~\ref{sec:polarized}.  
Thus, in discussion of many-body physics, there is no loss of generality 
in assuming that the microscopic structure of the magnetic dipole moment of a highly magnetic atom Eq.~(\ref{eq:magDoperator}) 
may be replaced by the effective structure Eq.~(\ref{eq:mutoF}).

\section{Appendix C: The mapping from symmetric top molecules to magnetic dipoles in second quantization}

In this section, we demonstrate that there is a correspondence 
between the second-quantized many-body Hamiltonians for symmetric top molecules and those for magnetic dipoles.  
We begin with the second quantized Hamiltonian for spin-$F$ magnetic dipoles~\cite{Ueda_2010}, 
\beq
\label{eq:secquantMag}
\hat{H}
&=&
\sum_{M_F=-F}^{F}
\int{\rm d}{\bf r}_1\:\psi_{FM_F}^{\dagger}({\bf r}_1)
\hat{H}^{(m)}_1
\psi_{FM_F}({\bf r}_1)
\\
&&+
\frac12\int{\rm d}{\bf r}_1{\rm d}{\bf r}_2
\sum_{i,j=X,Y,Z} \!\!
:
\hat{M}^F_{i}({\bf r}_1)
\frac{
\mu_0Q_{ij}({\bf r_1-\mathbf{r}_2})
}{4\pi}
\hat{M}^F_{j}({\bf r}_2)
:
\:,~
\nonumber
\eeq
where the colons denote normal ordering and $\hat{\psi}^{\dagger}_{FM_F}(\mathbf{r})$ creates a magnetic dipole at position ${\bf r}$ 
with its internal state $|FM_F\rangle$, discussed in Sec.~\ref{sec:magdipoles}.  The field operator $\hat{\psi}^{\dagger}_{FM_F}(\mathbf{r})$ satisfies commutation or anti-commutation relations depending on the statistics of the magnetic dipoles.  Here, we have defined the single-particle Hamiltonian of the magnetic dipoles $\hat{H}^{(m)}_1$, 
which is composed of the kinetic term and the Zeeman term for ${\bf B}=B_{Z}{\bf e}_Z$,  Eq.~(\ref{eq:FOZE}), 
\beq
\hat{H}^{(m)}_1
&=&
-\frac{\hat{\bf p}^2}{2m}
-g_F \mu_B M_F  B_{Z}
~, 
\eeq
and the function $Q_{ij}$, the kernel of the dipole-dipole interaction, as
$Q_{ij}({\bf r})=[\delta_{ij}-3({\bf e}_{r})_i({\bf e}_{r})_j]/r^3$.  
We have also introduced local spin operators $\hat{M}^F_{i}({\bf r})$, 
\beq
\hat{M}^F_{i}({\bf r})
&=&
\sum_{M'_F,M_F=-F}^{F}
\psi_{FM'_F}^{\dagger}({\bf r})
(\hat{f}_{i})_{FM'_F,FM_F}
\psi_{FM_F}({\bf r})
~,
\label{eq:localmag}
\eeq
with the matrix elements of the magnetic dipole moment operator 
$(\hat{f}_{i})_{FM'_F,FM_F}$ $(i=X,Y,Z)$ defined by
\beq
(\hat{f}_{i})_{FM'_F,FM_F}
&=&
\langle FM'_{F'}|\hat{\mu}_{i}|FM_F\rangle
~.
\label{eq:matelMD}
\eeq
The explicit form of Eq.~(\ref{eq:matelMD}) is given by Eq.~(\ref{eq:totalMD}) in Sec.~\ref{sec:mapping}. 

On the other hand, the second quantized Hamiltonian for symmetric top molecules is given by 
\beq
\hat{H}
&=&
\label{eq:secquantEle}
\sum_{JKM}
\int{\rm d}{\bf r}_1\:\psi_{JKM}^{\dagger}({\bf r}_1)
\hat{H}^{(e)}_1
\psi_{JKM}({\bf r}_1)
\\
&&+\frac12\int{\rm d}{\bf r}_1{\rm d}{\bf r}_2
\sum_{i,j=X,Y,Z} \!\!
:
\hat{\cal{D}}_{i}({\bf r}_1)
\frac{
Q_{ij}({\bf r_1-\mathbf{r}_2})
}{4\pi\varepsilon_0}
\hat{\cal{D}}_{j}({\bf r}_2)
:
\:,~
\eeq
where $\hat{\psi}^{\dagger}_{JKM}(\mathbf{r})$ creates a symmetric top molecule at position ${\bf r}$ 
with its internal state $|JKM\rangle$, discussed in Sec.~\ref{sec:ST}, 
and satisfies commutation or anti-commutation relations depending on the statistics of the molecules. 
Here, $\hat{H}^{(e)}_1$ denotes the single-particle Hamiltonian of symmetric top molecules, 
composed of the kinetic term and the linear Stark term for ${\bf E}=E_{\rm DC}{\bf e}_Z$, Eq.~(\ref{eq:FOSE}),
\beq
\hat{H}^{(e)}_1
&=&
-\frac{\hat{\bf p}^2}{2m}
-\frac{dK}{J(J+1)}M E_{\mathrm{DC}}
~. 
\eeq
Similar to the local magnetization operators, Eq.~(\ref{eq:localmag}), 
local dipole moment operators $\hat{\cal{D}}_{i}({\bf r})$ are defined by
\beq
\hat{\cal D}_{i}({\bf r})
&=&
\sum_{J'\!K'\!M'\!,JKM}
\psi_{J'\!K'\!M'}^{\dagger}({\bf r})
(\hat{d}_{i})_{{J'\!K'\!M'\!,JKM}}
\psi_{JKM}({\bf r})
~,
\eeq
with the matrix elements of the electric dipole moment operator for symmetric top molecules 
$(\hat{d}_{i})_{J'\!K'\!M'\!,JKM}$ $(i=X,Y,Z)$, 
\beq
(\hat{d}_{i})_{J'\!K'\!M'\!,JKM}
&=&
\bra{J'K'M'}\hat{d}_i\ket{JKM}
~.
\eeq 
We remark that the summation of $J,K,M$ runs over all non-negative integers for $J$, 
and integers such that $-J\leq K\leq J$, $-J\leq M\leq J$ for $K$ and $M$.
As discussed in Sec.~\ref{sec:ST}, for the symmetric top molecules in a fixed $(J,K)$-manifold, we have
\beq
(\hat{d}_p)_{JKM',JKM}
&=&
(-1)^{J-M'}
\begin{pmatrix}
J & 1 & J \\
-M' & p & M
\end{pmatrix}
\langle JK'|\!|\hat{\mathbf{d}}|\!|JK\rangle
~,
\eeq
with the reduced matrix element given by Eq.~(\ref{eq:redmatelem} ). 
Thus, for any given species of magnetic dipoles with integral spin-$F$, 
we can choose the corresponding $(2F+1)$ states from rotational states of symmetric top molecules 
with $J=F$ and $K\neq0$ such that
\beq
(\hat{d}_i)_{FKM',FKM}
&=&
\frac{\langle JK'|\!|\hat{\mathbf{d}}|\!|JK\rangle}{g_F \mu_B \langle F|\!|\hat{\bf F}|\!|F\rangle }
\langle FM'_{F}|\hat{\mu}_{i}|FM_F\rangle
\nonumber\\
&=&
\frac{dK}{g_F \mu_B F(F+1)}
(\hat{f}_{i})_{Fm'_F,Fm_F}
~.
\eeq
When we concentrate on the fixed $(J,K)$-manifold of the Hamiltonian Eq.~(\ref{eq:secquantEle})
with $J=F$ and $K\neq0$, it reduces to 
\beq
\label{eq:secquantEleFK}
\hat{H}
&=&
\sum_{M=-F}^{F}
\int{\rm d}{\bf r}_1\:\psi_{FKM}^{\dagger}({\bf r}_1)
\hat{H}^{(e)}_1
\psi_{FKM}({\bf r}_1)\\
\nonumber &&+
\frac12\int{\rm d}{\bf r}_1{\rm d}{\bf r}_2
\sum_{i,j=X,Y,Z} \!\!
:
\hat{\cal{D}}^{FK}_{i}({\bf r}_1)
\frac{
Q_{ij}({\bf r_1-\mathbf{r}_2})
}{4\pi\varepsilon_0}
\hat{\cal{D}}^{FK}_{j}({\bf r}_2)
:
\:,~
\eeq
with local dipole moment operators in the $(F,K)$-manifold 
$\hat{\cal{D}}^{FK}_{i}({\bf r})$ given by, 
\beq
\hat{\cal D}^{FK}_{i}({\bf r})
&=&
\sum_{M'\!,M=-F}^{F}
\psi_{F\!K\!M'}^{\dagger}({\bf r})
(\hat{d}_{i})_{{F\!K\!M'\!,FKM}}
\psi_{FKM}({\bf r})
~. 
\eeq
Comparing Eq.~(\ref{eq:secquantEleFK}) with Eq.~(\ref{eq:secquantMag}), 
we conclude that the second quantized Hamiltonian for symmetric top molecules can simulate the many-body physics of magnetic dipoles.

\end{document}